# Thermal investigation of bistability in high index doped silica integrated ring resonators

*David J. Moss*

dmoss@swin.edu.au

Optical Sciences Center, Swinburne University of Technology, Hawthorn, VIC 3122, Australia

**Abstract**

The utilization and engineering of thermo-optic effects have found broad applications in integrated photonic devices, facilitating efficient light manipulation to achieve various functionalities. Here, we perform both an experimental characterization and theoretical analysis of these effects in integrated micro-ring resonators in high index doped silica (HIDS), which has had many applications in integrated photonics and nonlinear optics. By fitting the experimental results with theory, we obtain fundamental parameters that characterize their thermo-optic performance, including the thermo-optic coefficient, the efficiency for the optically induced thermo-optic process, and the thermal conductivity. The characteristics of these parameters are compared to those of other materials commonly used for integrated photonic platforms, such as silicon, silicon nitride, and silica. These results offer a comprehensive insight into the thermo-optic properties of HIDS based devices. Understanding these properties is essential for efficiently controlling and engineering them in many practical applications.

## I. INTRODUCTION

Heat management and control of optical devices is of fundamental importance for their practical applications [1, 2]. For integrated photonic devices with a compact footprint and tight mode confinement, and particularly for materials that do not exhibit second-order optical nonlinearities such as the Pockels effect [3], the importance of precisely engineering their thermo-optic effects is even more pronounced [4, 5]. Over the past decade, with the rapid advancement of integrated photonics, extensive research has been dedicated to investigating and harnessing thermo-optic effects to manipulate light in integrated photonic devices, particularly those based on centrosymmetric materials [4, 6]. This has enabled the realization of a variety of functionalities such as mode-locking [7, 8], optical switches [9, 10], logic gates [11], power limiters [11, 12], and optical memories [11, 13, 14].

As an important complementary metal–oxide–semiconductor (CMOS)-compatible integrated platform, high index doped silica (HIDS) has been extensively utilized for diverse linear and nonlinear integrated photonic devices for a range of applications [15-22]. HIDS possesses a host of attractive optical properties, such as low linear optical absorption over a broad band, a reasonably strong Kerr nonlinearity (about 5 times that of silica), and negligible nonlinear optical absorption [23-26]. The combination of these properties and its strong compatibility with the globally established CMOS infrastructure, contributes to the exceptional performance and versatility of HIDS devices in various applications within the field of integrated photonics.

Despite its proven success for many optical applications, investigation of the thermo-optic effects in HIDS devices has not been as extensive as in other integrated photonic devices made from other centrosymmetric materials such as silicon and silicon nitride [14, 27, 28]. There remains a need for exploration and an understanding of the thermo-optic properties of HIDS devices to fully leverage their potential in integrated photonics. In this paper, we address this issue by providing a comprehensive experimental characterization and theoretical analysis of these effects in HIDS integrated devices. By fitting experimental results with theory, we obtain fundamental parameters that characterize the thermo-optic properties of HIDS devices, including the thermo-optic coefficient, the efficiency for the optically induced thermo-optic process, and the thermal conductivity. We also provide a comparison of these parameters with those of other materials used for CMOS-compatible integrated photonic platforms, such as silicon, silicon nitride, and silica. These findings provide a comprehensive understanding of the thermo-optic properties of HIDS devices, important for effectively controlling and engineering these devices in many applications.

## II.     DEVICE FABRICATION AND CHARACTERISATION

**Fig. 1(a)** shows a schematic of an add-drop MRR made from HIDS. A microscope image of the fabricated device is shown in **Fig. 1(b)**. The MRR was fabricated via CMOS-compatible processes [15, 18]. First, the HIDS film with a refractive index of ~1.66 was deposited using plasma enhanced chemical vapor deposition (PECVD). Next, waveguides with exceptionally low sidewall roughness were formed by employing deep ultraviolet photolithography techniques and reactive ion etching. Finally, a silica layer with a refractive index of ~1.45 was deposited via PECVD as the upper cladding. The waveguide cross section of both the MRR and the two coupling bus waveguides was ~3 μm × ~2 μm. The MRR had a radius of ~592.1 μm, which corresponded to a free spectral range (FSR) of ~0.4 nm (*i.e.*, ~49 GHz). Note that although there were a number of concentric rings in **Fig. 1(b)**, only the central ring was coupled with the through / drop bus waveguides to form an MRR with a radius of ~592.1 μm – the rest were simply used to enable easy identification by eye. A similar MRR layout was used in our previous work on HIDS devices [25, 26]. The input and output ports of the MRR were connected to specially designed mode converters that were packaged with fiber pigtails. The fiber-to-chip coupling loss was ~1.5 dB / facet, with this low value enabled through the use of on-chip mode converters to the pigtailed fibers.

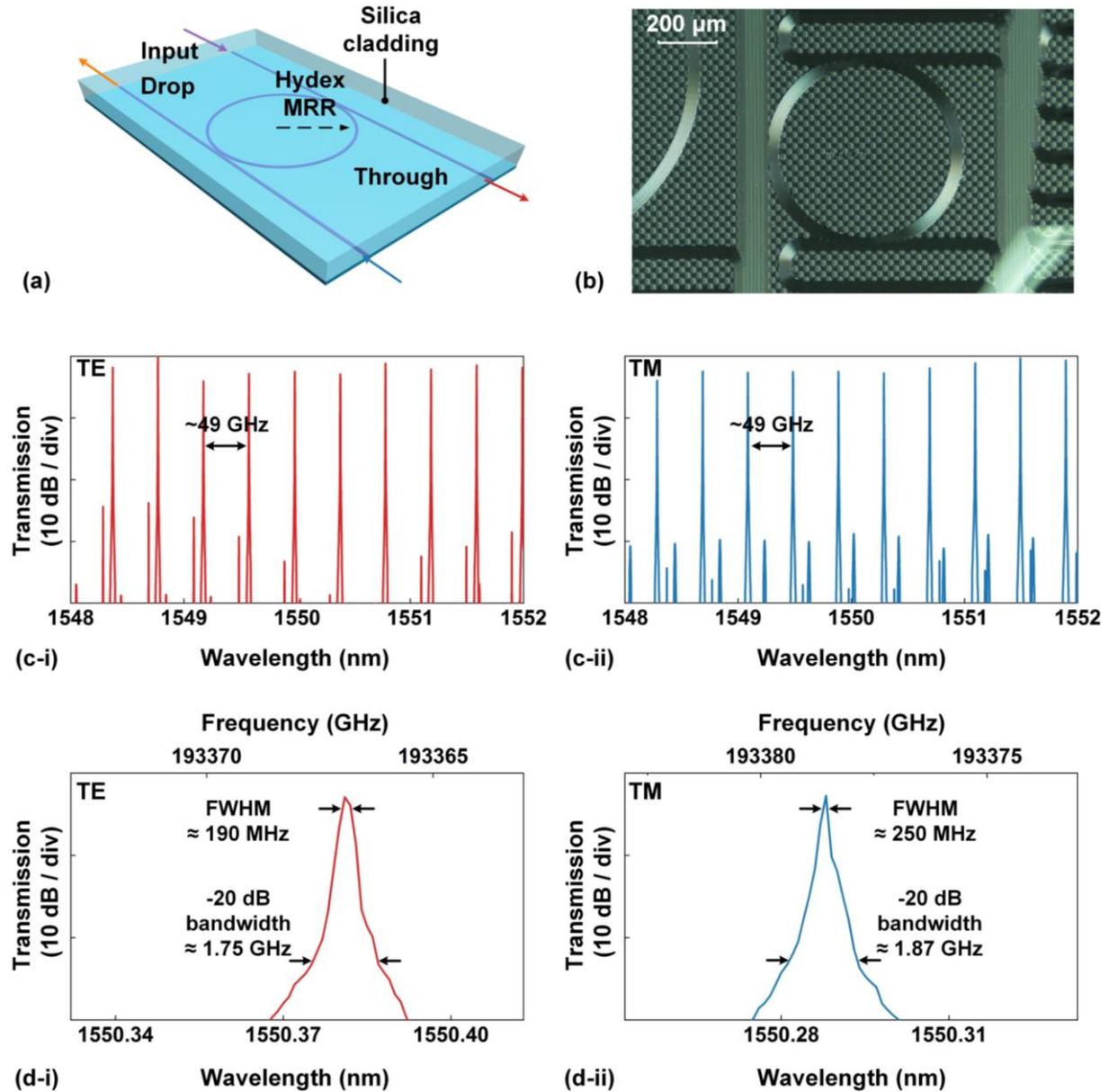

Fig. 1. (a) – (b) Schematic and microscope image of an add-drop microring resonator (MRR) made from high index doped silica (HIDS), respectively. (c) Measured transmission spectra of the HIDS MRR for (i) TE and (ii) TM polarizations. (d) Zoom-in views of single (i) TE- and (ii) TM- polarized resonances at ~1550.381 nm and ~1550.288 nm, respectively.

**Fig. 1(c)** shows the measured transmission spectra of a fabricated HIDS MRR for both transverse magnetic (TE) and transverse electric (TM) polarizations. The wavelength of a tunable continuous-wave (CW) laser was scanned at a constant input power of ~0 dBm to measure the transmission spectra, and a polarization controller (PC) was employed to adjust the input polarization. The input power here and in our following analysis refers to the power coupled into the device (*i.e.*, the on-chip power), with the fiber-to-chip coupling loss being subtracted from the laser's output power. The free spectral range (FSR) of the TE- and TM- polarized transmission spectra was ~0.4 nm, which corresponded to ~49 GHz. By tuning the PC, the maximum polarization extinction ratios for the TE- and TM- polarized resonances were > 30 dB.

TABLE I. DEVICE PARAMETERS OF THE HIDS MRR

| | Parameter | Symbol | Value | Source |
|---|---|---|---|---|
| Material parameters | Refractive index | $n$ | silica: 1.45<br>HIDS: 1.66 | [18, 29] |
| | Electrical conductivity (S / m) | $\sigma$ [a] | $1.0 \times 10^{-10}$ | [15] |
| Waveguide parameters | Width (μm) | $W$ | ~3 | Device structural parameter |
| | Height (μm) | $H$ | ~2 | Device structural parameter |
| MRR parameters | Ring radius (μm) | $R$ | ~592.1 | Device structural parameter |
| | Field transmission coefficients | $t_{1,2}$ [b] | TE: 0.9985<br>TM: 0.9980 | Fit results from Fig.1(d) |
| | Round-trip amplitude transmission | $a$ | TE: 0.9970<br>TM: 0.9976 | Fit results from Fig. 1(d) |
| | Intensity build-up factor | $BUF$ | TE: ~11.2<br>TM: ~11.4 | Calculated based on the fitted $t_{1,2}$ and $a$ |

a) $\sigma$ is the electrical conductivity of silica and HIDS. Here we neglect the difference between the electrical conductivities of silica and HIDS since both of them are dielectrics with extremely low electrical conductivities.
b) The field transmission coefficients of the two couplers formed by the MRR and the two bus waveguides are assumed to be equal, *i.e.*, $t_1 = t_2$.

**Fig. 1(d)** shows zoom-in views of single TE- and TM- polarized resonances at ~1550.381 nm and ~1550.288 nm. There was no significant asymmetry in the measured resonance spectral lineshape, indicating that the thermal effect at the input power of ~0 dBm was negligible. The full widths at half maximum (FWHMs) of the TE- and TM-polarized resonances were ~0.0015 nm (~190 MHz) and ~0.0020 nm (~250 MHz), respectively, which corresponded to Q factors of ~1.0 × $10^6$ and ~7.8 × $10^5$, respectively. In addition, the -20-dB bandwidths of the TE- and TM- polarized resonances were ~1.75 GHz and ~1.87 GHz, respectively. By using the scattering matrix method [30, 31] to fit the measured spectra in **Fig. 1(d)**, we obtained the device parameters for the HIDS MRR that will be used for analysis in the subsequent sections. These parameters, together with the specific material and waveguide parameters, are summarized in **Table I**.

### III. THERMO-OPTIC COEFFICIENT

The thermo-optic coefficient of a material is a fundamental parameter that indicates how its refractive index changes with environmental temperature, which plays an important role in the design and engineering of relevant devices [32]. In this section, we characterize the thermo-optic coefficient of HIDS by measuring the transmission spectra of the HIDS MRR with varying chip temperature.

When there are changes in environmental temperature, the thermo-optic effect causes changes in the effective refractive index of the HIDS waveguides. Consequently, this leads to a shift in the resonance wavelengths of the HIDS MRR. **Fig. 2(a)** shows the TE- and TM-polarized transmission spectra of the HIDS MRR when the chip temperature changed from 23 °C to 30 °C, respectively. We measured the shifts of three resonances, including a TE-polarized resonance and two TM-polarized resonances (TM1 and TM2). Specially, the TE-polarized resonance was located between the two TM-polarized resonances. To adjust the temperature of the integrated chip mounted on a stage, a temperature controller was employed. The input power of the scanned CW laser was maintained as ~0 dBm (*i.e.*, the same as that in **Fig. 1**) in order to mitigate noticeable thermal effects. It is important to highlight that despite the changes in environmental temperature, no significant asymmetry was observed in the measured resonance spectral lineshape. This observation indicates that changes in environmental temperature induced by the temperature controller have a minimal impact on the asymmetry of the resonance spectral lineshape and does not induce significant optical bistability that will be discussed in the next section [33].

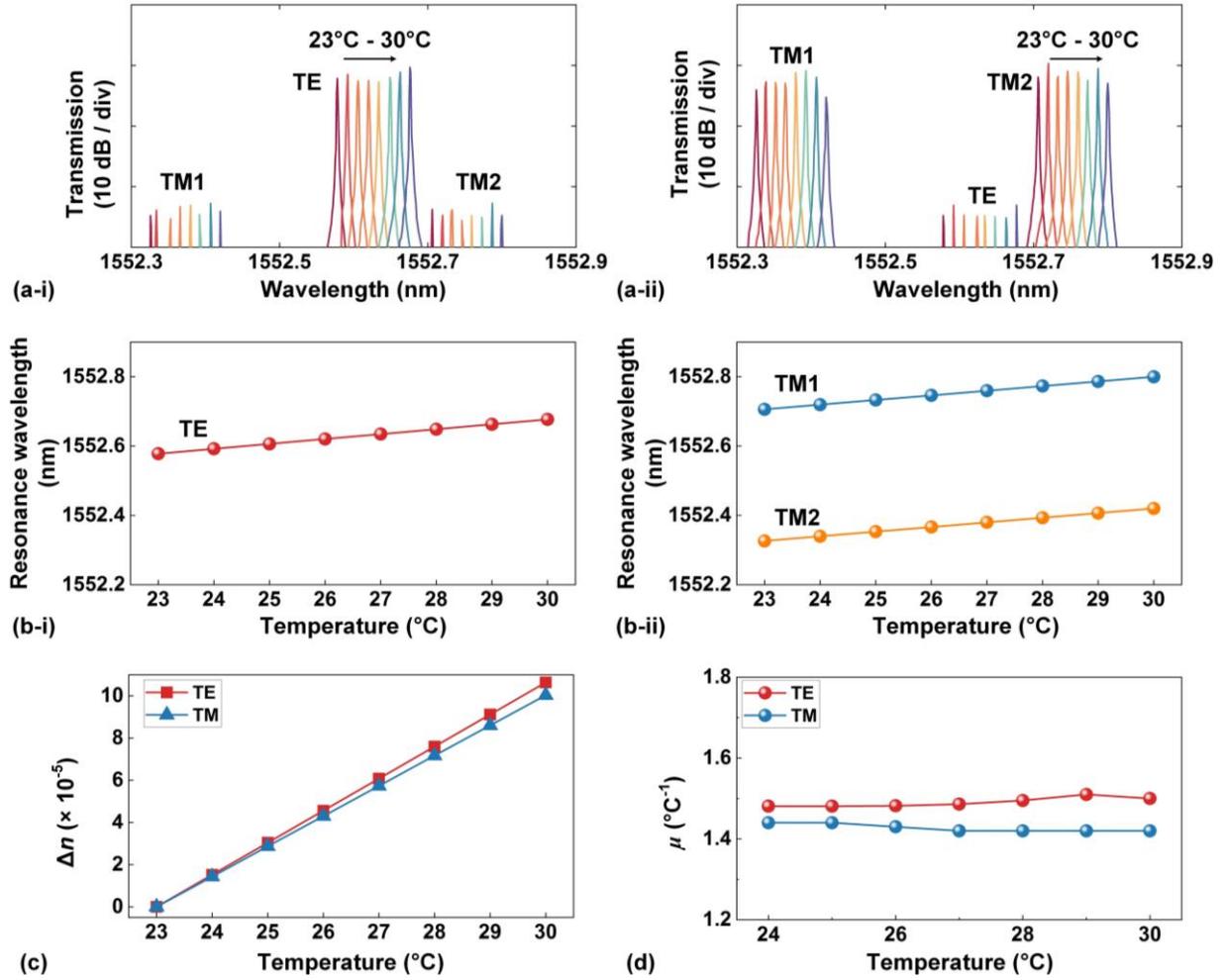

Fig. 2. (a) Measured (i) TE- and (ii) TM- polarized transmission spectra of the HIDS MRR when the chip temperature changes from 23 °C to 30 °C, respectively. The results presented depict a resonance with TE polarization positioned between two resonances with TM polarization (TM1 and TM2). (b) Resonance wavelength shifts versus chip temperature for (i)TE and (ii) TM polarizations extracted from (a). (c) Changes in waveguide effective refractive indices versus chip temperature extracted from (b). (d) Thermo-optic coefficient $\mu$ versus chip temperature extracted from (c).

**Fig. 2(b)** shows the resonance wavelength shifts versus the chip temperature, which were extracted from the results in **Fig. 2(a)**. The TE-polarized resonance redshifted at a rate of ~14.2 pm / °C, whereas the two TM-polarized resonances exhibited a redshift rate of ~13.4 pm / °C. **Fig. 2(c)** depicts the changes in the waveguide effective refractive indices versus the chip temperature. These results were calculated using the measured results in **Fig. 2(b)**, along with the relationship between the resonance wavelengths and the waveguide effective refractive index given by [34, 35]:

$$n_{eff} \cdot 2\pi / \lambda_m \cdot L = m \cdot 2\pi, \tag{1}$$

where $n_{eff}$ is the effective refractive index of the HIDS waveguide, $L$ is the circumference of the

HIDS MRR, and *m* represents the *m*th resonance, with $\lambda_m$ denoting the corresponding resonance wavelength.

In **Fig. 2(c)**, the TE mode displays a change in the effective refractive index at a rate of ~$1.52 \times 10^{-5}$ /°C, while the TM mode changes at a rate of ~$1.43 \times 10^{-5}$ /°C. The difference in these rates can be attributed to the asymmetric cross section of the HIDS waveguide. Based on these results, we further extract the thermo-optic coefficient of the HIDS material at various chip temperatures by using Lumerical FDTD commercial mode solving software. The results are presented in **Fig. 2(d)**. In our simulation, the thermo-optic coefficient of silica is assumed to be ~$1.09 \times 10^{-5}$ /°C [36]. The thermo-optic coefficients of HIDS in **Fig. 2(d)** do not show significant temperature dependence. We also note that the average values of the thermo-optic coefficients of HIDS derived from the TE and TM modes exhibit remarkable similarity, at ~$1.49 \times 10^{-5}$ /°C and ~$1.44 \times 10^{-5}$ /°C, respectively. This close resemblance between the coefficients reflects that the HIDS does not exhibit significant anisotropy in terms of its thermo-optic coefficient.

## IV. OPTICALLY INDUCED THERMO-OPTIC RESPONSE

When a material is illuminated with intense light, optical absorption leads to heat generation that raises the local temperature. This in turn modifies the material's refractive index, thereby influencing the propagation of light through the material. In this optically induced thermo-optic process, the change in the material's refractive index *n* due to the temperature variation induced by the optical field can be modeled as [27, 37]:

$$n = n_0 + \bar{n}_2 \cdot I, \qquad (2)$$

where $n_0$ is the material's refractive index when not exposed to light, and $\bar{n}_2 \cdot I$ is the refractive index change due to the optically induced temperature change, with *I* denoting the light intensity and $\bar{n}_2$ denoting the coefficient that characterizes the efficiency for this process. In this section, we characterize the $\bar{n}_2$ of HIDS by measuring the transmission spectra of the HIDS MRR at various input powers. It is worth noting that **Eq. (2)** is the same as that used for modeling the nonlinear Kerr optical effect [38, 39]. We also note that $\bar{n}_2$ is an effective response, in that it is device geometry dependent, including the MRR Q-factor, coupling coefficient etc..

For the optically induced refractive index change, in addition to the optically induced thermo-optic effect, there will also be a presence of the Kerr optical effect. Despite having the same mathematical modeling as shown in **Eq. (2)**, these two effects are associated with different physical processes that exhibit distinct characteristics. For example, compared to the Kerr optical effect that has an ultrafast time response on the order of $10^{-15}$ s [40, 41], the time response for the optically induced thermo-optic effect is much slower, typically on the order of $10^{-6} - 10^{-3}$ s [31, 42].

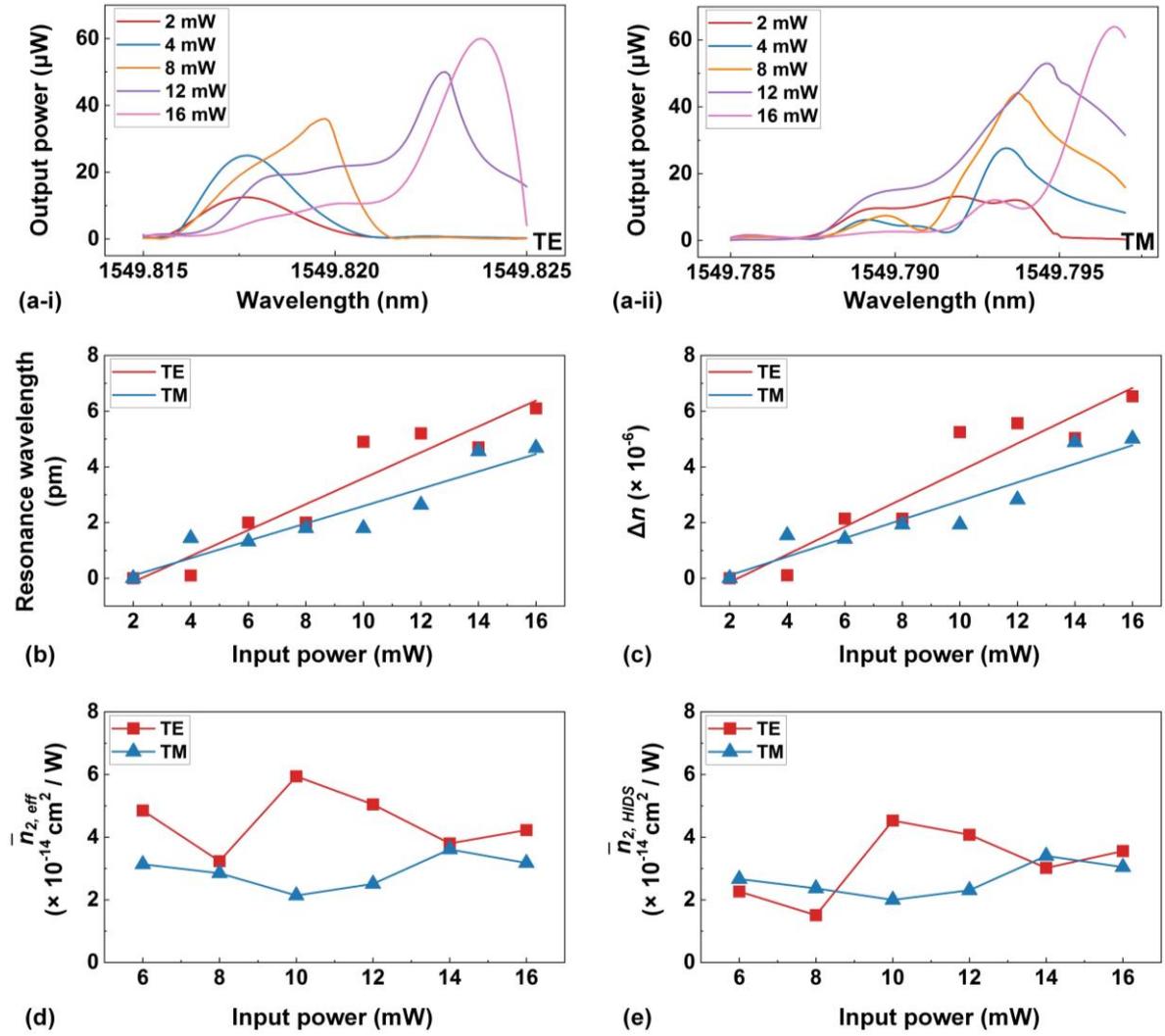

Fig. 3. (a) Measured transmission spectra of the HIDS MRR at varying input powers for (i) TE and (ii) TM modes. (b) Measured (data points) and fitted (solid curves) resonance wavelength shifts versus input power. (c) Waveguide effective refractive index changes versus input power extracted from (b). (d) $\bar{n}_{2,\,eff}$ versus input power extracted from (c). (e) $\bar{n}_{2,\,HIDS}$ versus input power extracted from (d).

When the wavelength of incident light is on resonance with the MRR, the incident light power converts into heat more efficiently, being enhanced significantly by the ring resonance, leading to an efficient change in the effective refractive index of the HIDS waveguides caused by the thermo-optic effect. This refractive index change also results in a shift in the resonance wavelengths of the HIDS MRR. **Figs. 3(a-i)** and **(a-ii)** show measured transmission spectra of the HIDS MRR at different input powers for TE and TM polarizations, respectively. As the input power increased, a redshift in the resonance wavelengths was observed, accompanied by increasingly asymmetric resonance spectra. The spectra also exhibit a steepened transition edge, indicating the presence of the optical bistability [43, 44].

Depending on the dominating nonlinear mechanism, the resonance wavelengths can experience either a blue or red shift. In previous work on bistability in silicon MRRs at room temperature, it was observed that the resonance wavelengths initially exhibited a blueshift and subsequently transitioned to a redshift as the input power increased [45]. This is because the free-carrier dispersion (FCD) that results in a decreased refractive index of silicon dominates at low powers, whereas the thermo-optic effect that leads to an increased refractive index dominates at high powers [45]. Here, we only observed a redshift in the resonance wavelengths, mainly due to the dominating thermo-optic effect and negligible FCD for the HIDS MRR [26, 29] and the fact that the magnitude of the all-optical Kerr component of the index change tends to be much smaller for typical CW powers.

**Fig. 3(b)** shows the shifts of the resonance wavelength versus the input power. For both the TE- and TM- polarizations, the positive $\Delta\lambda$ (which indicates a redshift) exhibits a nearly linear relationship with the input power. By linearly fitting the measured results, we obtained the rates for the resonance wavelength shift, which were ~0.4655 pm / mW and ~0.3144 pm / mW for the TE and TM polarizations, respectively.

**Fig. 3(c)** shows the changes in the waveguide effective refractive indices versus the input power for both TE and TM polarizations. These results were calculated based on **Eq. (1)**, using the measured results in **Fig. 3(b)**. As the input power increased from ~2 mW to ~16 mW, the effective refractive indices of TE and TM modes displayed changes of ~6.533 × $10^{-6}$ and ~5.012 × $10^{-6}$, respectively. These changes correspond to average rates of ~4.985 × $10^{-7}$ /mW and ~3.335 × $10^{-7}$ /mW, respectively.

**Fig. 3(d)** shows the waveguide effective $\bar{n}_2$, denoted as $n_{2,\text{eff}}$, versus the input power for both TE and TM polarizations, which were extracted from the results in **Fig. 3(c)**. The $\bar{n}_{2,\text{eff}}$ was calculated by [27]

$$\bar{n}_{2,\,eff} = \Delta n / I, \tag{3}$$

where $\Delta n$ is the refractive index change, and $I$ is the light intensity in the MRR given by [27]

$$I = \frac{P_{in} \cdot BUF}{A_{eff}}. \tag{4}$$

In **Eq. (4)**, $P_{in}$ is the input power, $A_{eff}$ is the effective mode area [26, 46], and $BUF$ is the intensity build-up factor of the MRR that can be expressed [47, 48]

$$BUF = \frac{(1 - t_1^2)t_2^2 a^2}{1 - 2t_1 t_2 a + (t_1 t_2 a)^2}, \tag{5}$$

where $t_{1,2}$ and $a$ are the fit MRR parameters in **Table 1**.

In **Fig. 3(d)**, the average values of the extracted $\bar{n}_{2,\,eff}$ for the TE and TM polarizations are ~4.514 × 10$^{-14}$ cm$^2$ / W and ~2.903 × 10$^{-14}$ cm$^2$ / W, respectively. The difference in these responses can be attributed to the asymmetric cross section of the HIDS waveguide that results in different optical field distributions for the two modes. Based on the results in **Fig. 3(d)**, we further extract the $\bar{n}_2$ for the HIDS material, denoted as $\bar{n}_{2,\,HIDS}$, according to [26, 49]:

$$\bar{n}_{2,\,eff} = \frac{\iint_D n_0^2(x,y) \bar{n}_2(x,y) S_z^2 dxdy}{\iint_D n_0^2(x,y) S_z^2 dxdy} \tag{6}$$

where $D$ is the integral of the optical fields over the material regions, $S_z$ is the time-averaged Poynting vector calculated using Lumerical FDTD commercial mode solving software, $n_0(x, y)$ and $\bar{n}_2(x, y)$ are the linear refractive index and $\bar{n}_2$ profiles over the waveguide cross section, respectively. The value of $\bar{n}_2$ for silica used in our calculation was ~1.8 × 10$^{-14}$ cm$^2$ / W [36, 50]. **Fig. 3(e)** shows the extracted $n_{2,\,HIDS}$ versus the input power. The average values of $\bar{n}_{2,\,HIDS}$ derived from the TE and TM modes are ~3.1 × 10$^{-14}$ cm$^2$ / W and ~2.7 × 10$^{-14}$ cm$^2$ / W, respectively. The close resemblance between them reflects that the HIDS does not exhibit significant anisotropy in terms of its $\bar{n}_2$. The results in **Fig. 3(e)** also confirm that the predominant cause of the observed nonlinearity is thermal in nature. This is also supported by the fact that the Kerr nonlinear coefficient of HIDS (~1.3 × 10$^{-15}$ cm$^2$ / W [18, 51]) was over one order of magnitude lower. Although there are minor fluctuations in $\bar{n}_{2,\,HIDS}$ across various input powers in **Fig. 3(d)**, these variations are not significant. Considering the limited input power range (*i.e.*, ~2 mW to ~16 mW), it can be inferred that the $\bar{n}_{2,\,HIDS}$ values will exhibit a relatively stable behavior [15, 18]. Hence, the slight power-dependent variations in $\bar{n}_{2,\,HIDS}$ are likely attributable to measurement errors.

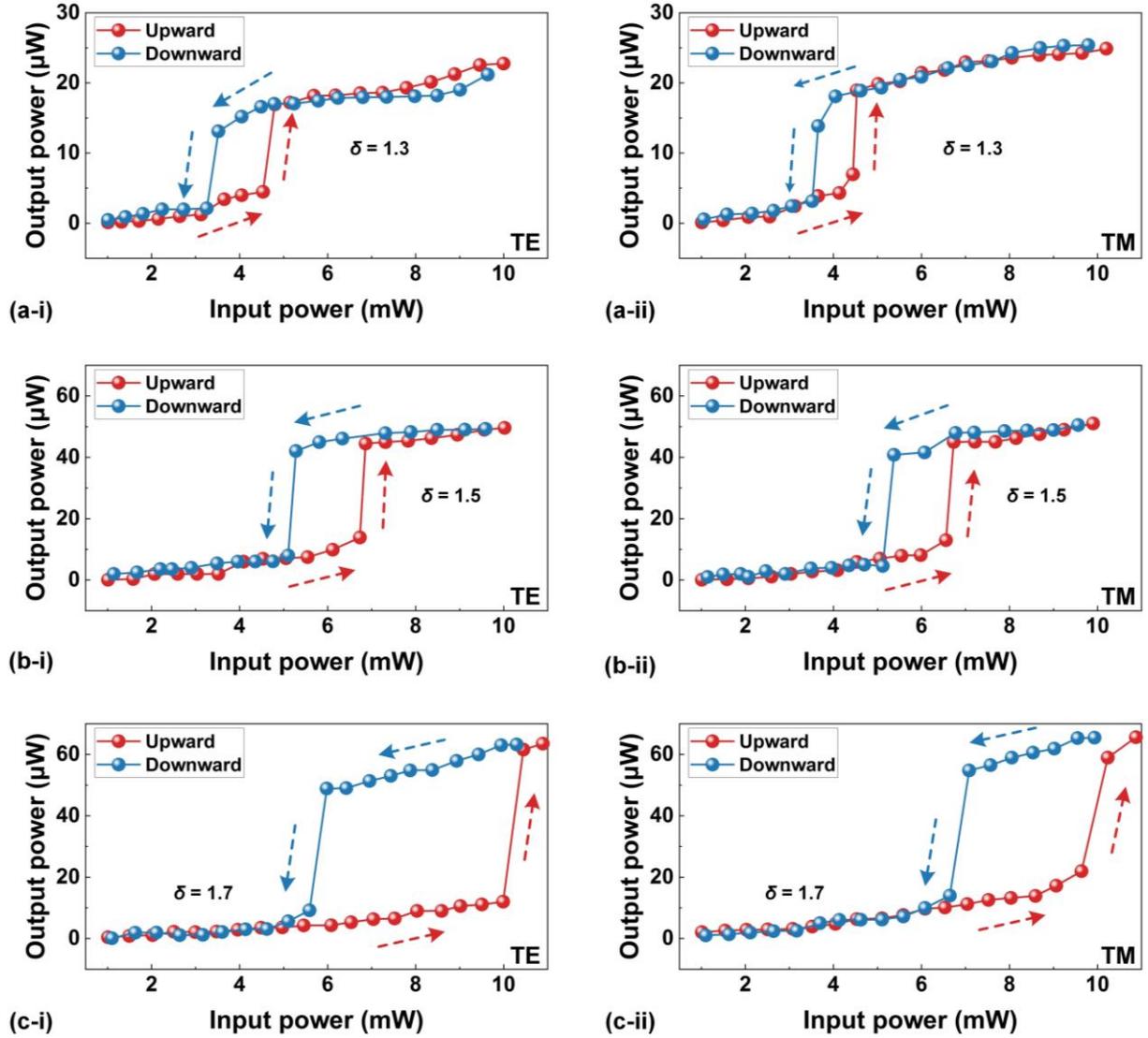

Fig. 4. Measured output power versus input power with initial wavelength detunings of (a) $\delta = \sim 1.3$, (b) $\delta = \sim 1.5$, and (c) $\delta = \sim 1.7$. In (a) – (c), (i) and (ii) show the results for TE- and TM- polarized resonances centered at ~1550.3758 nm and ~1550.2826 nm, respectively. Point-by-point measurements were taken at an average rate of ~1 Hz.

## V. OPTICAL BISTABILITY

Due to a steepened asymmetric transitional edge, optical bistability arising from nonlinear thermo-optic effects has been used for controlling light with light and achieving optical switches [9, 10]. **Fig. 4** shows the measured output power as a function of the input power when it was progressively increased from ~1 mW to ~8 mW. For comparison, we also plot the downward output power as the input power was subsequently reduced back to ~1 mW. In **Figs. 4(a) – (c)**, we show the results for three initial wavelength detunings of $\delta = \sim 1.3, \sim 1.5,$ and $\sim 1.7$, respectively. The $\delta$ is defined as:

$$\delta = (\lambda_{laser} - \lambda_{res}) / \Delta\lambda, \tag{7}$$

where $\lambda_{laser}$ is the wavelength of the input CW light, $\lambda_{res}$ is the resonance wavelength of the MRR measured at a low input CW power of ~0 dBm (*i.e.*, the same as that in **Fig. 1** and does not induce significant asymmetry in the measured resonance spectral lineshape), and $\Delta\lambda$ is the 3-dB bandwidth of the resonance. In our measurements, we chose a TE-polarized resonance centered at $\lambda_{res}$ = ~1550.3758 nm and a TM-polarized resonance centered at $\lambda_{res}$ = ~1550.2826 nm. During the measurements, the maximum polarization extinction ratios were kept > 30 dB.

In **Figs. 4(a) – (c)**, redshifts of the resonance wavelengths can be observed for both TE and TM polarizations. During the upward sweeping, the output power first exhibited a steady and continuous increase, followed by a sudden jump towards higher output power. Conversely, during the downward sweeping with decreasing input power, there was a sudden jump toward lower output power after a gradual decrease in the output power. Clearly, the presence of a hysteresis loop resulting from the upward and downward wavelength sweeping provides evidence for the existence of optical bistability in the HIDS MRR [52]. As $\delta$ was increased from ~1.3 to ~1.7, the input power threshold for optical bistability increased, and the hysteresis loop became more open. These phenomena are similar to those observed in Refs. [53, 54]. We also note that the TE-polarized resonance exhibits a more open hysteresis loop compared with the TM-polarized resonance at the same $\delta$. This observation shows agreement with the relatively large redshift of the resonance wavelength for the TE polarization in **Fig. 3(b)**.

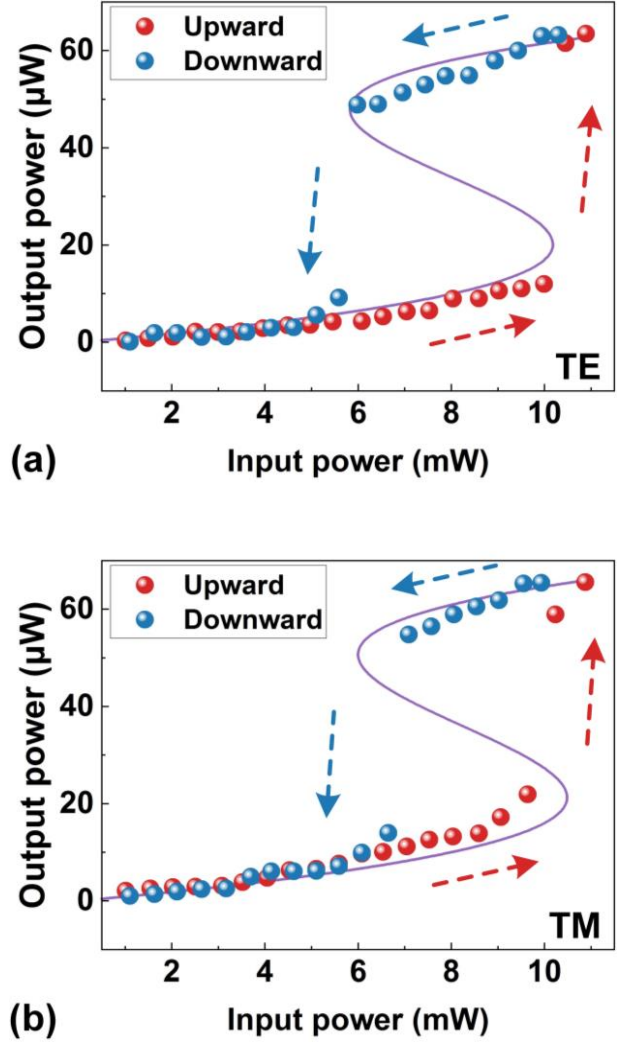

Fig. 5. Measured (data points) and theoretical (solid curves) output power versus input power for (a) TE and (b) TM polarizations. The initial wavelength detuning $\delta$ is ~1.7.

**Fig. 5** shows the measured and theoretical output powers versus the input power for both TE and TM polarizations. The theoretical curves were calculated based on the theory in Refs. [33, 52, 55], using both the device parameters in **Table I** and the fit $\bar{n}_{2,\,HIDS}$ in **Fig. 3(e)**. In principle, bistable behavior occurs in the resonator response because, under specific conditions, the output power yields multiple distinct solutions for a given input power. Consequently, the resonator can switch between these solutions due to the influence of noise [33, 55]. In **Fig. 5**, the measured results show good agreement with the theoretical curves, providing further confirmation of the accuracy of the fit thermo-optic property parameters for the HIDS devices.

## VI. THERMAL CONDUCTIVITY

Thermal conductivity, a parameter that defines a material's ability to conduct heat, has been widely used for modeling thermal transport for applications related to thermal management and energy storage [40, 56-60]. In this section, the thermal conductivity of HIDS is characterized by fitting the measured transmission spectra of the HIDS MRR at various input powers with theoretical simulations.

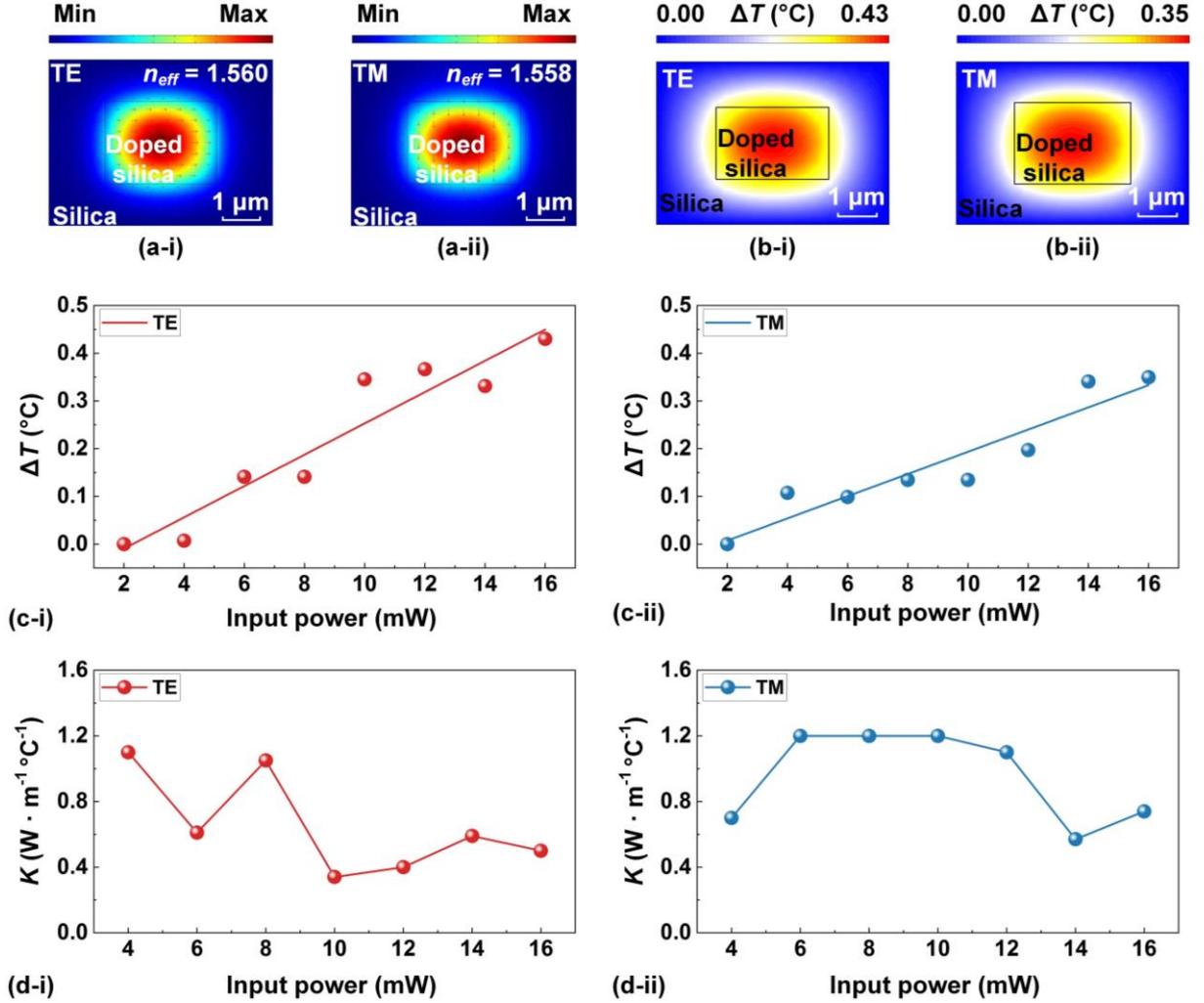

Fig. 6. (a) Optical mode profiles of the HIDS waveguide for (i) TE and (ii) TM modes. (b) Temperature distribution profiles of the HIDS waveguide for (i) TE and (ii) TM modes. In (a)–(b), the input CW power is ~16 mW and the initial temperature is assumed to be at room temperature of 23 °C. (c) Calculated temperature variation versus input power for (i) TE and (ii) TM modes. (d) Thermal conductivity $K$ versus input power for (i) TE and (ii) TM modes.

**Figs. 6(a-i)** and **(a-ii)** show the simulated TE and TM mode profiles for the HIDS waveguide. The corresponding effective refractive indices were $n_{eff\_TE}$ = ~1.560 and $n_{eff\_TM}$ = ~1.558 at 1550 nm. To further investigate the heat generated in the HIDS waveguide, we simulated the cross-sectional temperature distribution for both TE and TM polarizations. **Figs. 6(b-i)** and **(b-ii)** show the steady-state temperature distributions at an incident power of 16 mW, which were obtained by solving the heat equation [1]:

$$-\nabla \cdot (K\nabla T) = q, \tag{8}$$

where $T$ is the steady-state temperature distribution, $K$ is the thermal conductivity, and $q$ is the heat source resulting from the absorption of light. In **Eq. (8)**, $\nabla T$ denotes the gradient of $T$, and $\nabla$ acting on the vector function $K\nabla T$ is the corresponding divergence operator. The heat source was calculated based on the TE and TM mode profiles in **Figs. 6(a-i)** and **(a-ii)** using [61]:

$$q = \frac{1}{2}\sigma |E|^2, \tag{9}$$

where $\sigma$ is the electrical conductivity of the waveguide in **Table I** and $E$ is the amplitude of the optical field simulated in **Fig. 6(a)**. It is worth noting that the build-up factor $BUF$ in **Eq. (5)** was taken into account when calculating the optical intensity in the MRR. In our simulation, the initial temperature $T_0$ was set to 23°C, which was the ambient temperature during the experiments.

When there are changes in the input power, the material conducts heat, leading to a rise in temperature and a redshift of the resonance wavelength. According to the results in **Figs. 2(c)** and **3(c)**, we calculated the device temperature variation versus the input power. As shown in **Fig. 6(c)**, at an input power of 16mW, the temperature variations for the TE and TM modes are $\Delta T$ = ~0.4298 °C and ~0.3495 °C, respectively. By fitting these temperature variations with the temperature distributions in **Fig. 6(b)**, we obtained the thermal conductivity for the HIDS, as shown in **Fig. 6(d)**. For the TE and TM polarizations, the average values for the fitted thermal conductivity were ~0.66 W / (m · °C) and ~0.96 W / (m · °C). We note that the thermal conductivity of the HIDS is lower than that of silica (*i.e.*, ~1.4 W / (m · °C) [61]). This can be attributed to the introduction of the doping material that slows down the lattice vibration coupling and the energy transfer. The specific values depend on the type and concentration of the doping element used, as well as the material's fabrication method and structure. In our fabrication, the HIDS was developed by using the high-index-contrast materials such as $Ta_2O_5$ and polymeric systems [62, 63]. We note that the thermal conductivity of $Ta_2O_5$ is much lower than that of silica (i.e., ~0.026 W / (m · °C) [64, 65]), which could be a reason for the relatively low thermal conductivity of HIDS. Another possible reason is that the thermal conductivity of silica films grown by PECVD (*i.e.,* ~1.0 W / (m · °C) [66]) is slightly lower than the previously mentioned value of silica. Based on **Eq. (8)**, the low thermal conductivity of the HIDS waveguide restricts heat propagation, leading to a higher concentration of thermal energy within the waveguide. Consequently, this amplifies the temperature increase, which, in turn, facilitates the attainment of more pronounced optical bistability.

VII.    COMPARISON WITH OTHER INTEGRATED PLATFORM MATERIALS

In this section, we present a summary of the thermo-optic property parameters of HIDS devices obtained in **Sections III – V**, together with a comparison of them with those exhibited by other materials used for CMOS-compatible integrated photonic platforms.

TABLE II. COMPARISON OF THERMO-OPTIC PROPERTY PARAMETERS OF HIDS AND OTHER INTEGRATED PLATFORM MATERIALS

| Parameter | Thermo-optic coefficient (°C$^{-1}$) | Coefficient for optically induced thermo-optic process (cm$^2$ / W) | Thermal conductivity (W · m$^{-1}$ °C$^{-1}$) | Refs. |
|---|---|---|---|---|
| silicon | ~1.8 × 10$^{-4}$ (~86 pm / °C)[a] | ~7.8 × 10$^{-11}$ | ~149 | [37, 67-69] |
| silicon nitride | ~2.6 × 10$^{-5}$ (~11 pm / °C)[a] | ~1.5 × 10$^{-15}$ | ~29 | [27, 70, 71] |
| silica | ~1.1 × 10$^{-5}$ (~15 pm / °C)[a] | ~1.8 × 10$^{-14}$ | ~1.4 | [36, 37, 50] |
| HIDS[b] | ~1.46 × 10$^{-5}$ (~13.8 pm / °C)[a] | ~2.9 × 10$^{-14}$ | ~0.8 | This work |

a) Here we also show the corresponding results for the wavelength shifts of resonators caused by temperature variation. Note that these results may vary based on the specific device used.
b) Here we show the average values of the results for the TE and TM polarizations obtained in Sections III – V.

As shown in **Table II**, the thermo-optic coefficient of HIDS is higher than that of silica, but lower than those of silicon nitride and silicon. This can be attributed to the moderate refractive index of HIDS among these materials. In terms of the coefficient characterizing the efficiency for the optically induced thermo-optic process, HIDS exhibits a value that is below that of silicon, yet it surpasses those of silica and silicon nitride. This highlights its capability for implementing high-performance nonlinear thermo-optic devices. For the thermal conductivity, HIDS displays the lowest value among these materials. This benefits its applications for thermal mode locking in optical microcomb generation [7, 55]. In the process of optical microcomb generation, the diminished thermal conductivity of HIDS introduces a slow thermal reaction that influences the steady-state dynamics of the intracavity power. This, in turn, leads to a gradual correlation between the cavity detuning and the pump power. Such characteristic decreases the rate of adjustment for power augmentation within the cavity in order to generate optical microcombs. As a result, it becomes feasible to achieve simple generation of stable soliton crystal microcombs through manual tuning of the pump laser [22-24]. These results have direct implications for optical microcombs realized in this platform [72-97] which will impact their classical [98-131] and quantum [132-144] applications as well as integrated novel photonic devices incorporating new 2D materials [145-199].

## VIII. CONCLUSION

In summary, we provide detailed experimental characterization and theoretical analysis of the thermo-optic effects in integrated HIDS devices that have been successfully applied in various linear and nonlinear optical applications. By fitting the experimental results with theory, we obtain fundamental parameters that define the thermo-optic performance of HIDS devices, including the thermo-optic coefficient, the efficiency for the optically induced thermo-optic process, and the thermal conductivity. We also compare these parameters with those of other materials used for CMOS-compatible integrated photonic platforms, such as silicon, silicon nitride, and silica. Our finding provides valuable insights into the thermo-optic properties of HIDS devices, which are crucial for effectively controlling and engineering these devices across diverse applications.

## Conflict of Interest

The authors state that there are no conflicts of interest

## References

[1] Y. Li, W. Li, T. Han, X. Zheng, J. Li, B. Li, S. Fan, and C.-W. Qiu, "Transforming heat transfer with thermal metamaterials and devices," *Nature Reviews Materials,* vol. 6, no. 6, pp. 488-507, 2021.

[2] H. Zhu, Q. Li, C. Zheng, Y. Hong, Z. Xu, H. Wang, W. Shen, S. Kaur, P. Ghosh, and M. Qiu, "High-temperature infrared camouflage with efficient thermal management," *Light: Science & Applications,* vol. 9, no. 1, pp. 60, 2020/04/14, 2020.

[3] C. Wang, M. Zhang, X. Chen, M. Bertrand, A. Shams-Ansari, S. Chandrasekhar, P. Winzer, and M. Lončar, "Integrated lithium niobate electro-optic modulators operating at CMOS-compatible voltages," *Nature,* vol. 562, no. 7725, pp. 101-104, 2018/10/01, 2018.

[4] K. Wu, Y. Wang, C. Qiu, and J. Chen, "Thermo-optic all-optical devices based on two-dimensional materials," *Photonics Research,* vol. 6, no. 10, pp. C22-C28, 2018/10/01, 2018.

[5] B. Desiatov, I. Goykhman, and U. Levy, "Direct Temperature Mapping of Nanoscale Plasmonic Devices," *Nano Letters,* vol. 14, no. 2, pp. 648-652, 2014/02/12, 2014.

[6] Q. Huang, H. Yu, Q. Zhang, Y. Li, W. Chen, Y. Wang, and J. Yang, "Thermally enhanced responsivity in an all-silicon optical power monitor based on defect-mediated absorption," *Photonics Research,* vol. 9, no. 11, pp. 2205, 2021.

[7] Y. Sun, J. Wu, M. Tan, X. Xu, Y. Li, R. Morandotti, A. Mitchell, and D. J. Moss, "Applications of optical microcombs," *Advances in Optics and Photonics,* vol. 15, no. 1, pp. 86, 2023.

[8] T. Herr, V. Brasch, J. D. Jost, C. Y. Wang, N. M. Kondratiev, M. L. Gorodetsky, and T. J. Kippenberg, "Temporal solitons in optical microresonators," *Nature Photonics,* vol. 8, no. 2, pp. 145-152, 2014/02/01, 2014.

[9] C. Qiu, C. Zhang, H. Zeng, and T. Guo, "High-Performance Graphene-on-Silicon Nitride All-Optical Switch Based on a Mach–Zehnder Interferometer," *Journal of Lightwave Technology,* vol. 39, no. 7, pp. 2099-2105, 2021.

[10] C. Qiu, Y. Yang, C. Li, Y. Wang, K. Wu, and J. Chen, "All-optical control of light on a graphene-on-silicon nitride chip using thermo-optic effect," *Scientific Reports,* vol. 7, no. 1, pp. 17046, 2017/12/06, 2017.

[11] P. W. Smith, and W. J. Tomlinson, "Bistable optical devices promise subpicosecond switching," *IEEE Spectrum,* vol. 18, pp. 26-33, June 01, 1981, 1981.

[12] M. Shirdel, and M. A. Mansouri-Birjandi, "All-optical bistable switching, hard-limiter and wavelength-controlled power source," *Frontiers of Optoelectronics,* vol. 9, no. 4, pp. 560-564, 2016/12/01, 2016.

[13] H. Gibbs, "Optical Bistability: Controlling Light with Light, Academic Press," *Inc.: Orlando, FL, USA*, 1985.

[14] V. R. Almeida, and M. Lipson, "Optical bistability on a silicon chip," *Optics Letters,* vol. 29, no. 20, pp. 2387-2389, 2004/10/15, 2004.

[15] M. Ferrera, L. Razzari, D. Duchesne, R. Morandotti, Z. Yang, M. Liscidini, J. E. Sipe, S. Chu, B. E. Little, and D. J. Moss, "Low-power continuous-wave nonlinear optics in doped silica glass integrated waveguide structures," *Nature Photonics,* vol. 2, no. 12, pp. 737-740, 2008.

[16] M. Ferrera, Y. Park, L. Razzari, B. E. Little, S. T. Chu, R. Morandotti, D. J. Moss, and J. Azaña, "On-

[16] chip CMOS-compatible all-optical integrator," *Nature Communications,* vol. 1, no. 1, pp. 29, 2010/06/15, 2010.

[17] L. Razzari, D. Duchesne, M. Ferrera, R. Morandotti, S. Chu, B. E. Little, and D. J. Moss, "CMOS-compatible integrated optical hyper-parametric oscillator," *Nature Photonics,* vol. 4, no. 1, pp. 41-45, 2010/01/01, 2010.

[18] D. J. Moss, R. Morandotti, A. L. Gaeta, and M. Lipson, "New CMOS-compatible platforms based on silicon nitride and Hydex for nonlinear optics," *Nature Photonics,* vol. 7, no. 8, pp. 597-607, 2013.

[19] H. Bao, A. Cooper, M. Rowley, L. Di Lauro, J. S. Totero Gongora, S. T. Chu, B. E. Little, G.-L. Oppo, R. Morandotti, D. J. Moss, B. Wetzel, M. Peccianti, and A. Pasquazi, "Laser cavity-soliton microcombs," *Nature Photonics,* vol. 13, no. 6, pp. 384-389, 2019.

[20] M. Kues, C. Reimer, P. Roztocki, L. R. Cortés, S. Sciara, B. Wetzel, Y. Zhang, A. Cino, S. T. Chu, B. E. Little, D. J. Moss, L. Caspani, J. Azaña, and R. Morandotti, "On-chip generation of high-dimensional entangled quantum states and their coherent control," *Nature,* vol. 546, no. 7660, pp. 622-626, 2017/06/01, 2017.

[21] C. Reimer, M. Kues, P. Roztocki, B. Wetzel, F. Grazioso, B. E. Little, S. T. Chu, T. Johnston, Y. Bromberg, L. Caspani, D. J. Moss, and R. Morandotti, "Generation of multiphoton entangled quantum states by means of integrated frequency combs," *Science,* vol. 351, no. 6278, pp. 1176-1180, 2016.

[22] M. Rowley, P.-H. Hanzard, A. Cutrona, H. Bao, S. T. Chu, B. E. Little, R. Morandotti, D. J. Moss, G.-L. Oppo, J. S. Totero Gongora, M. Peccianti, and A. Pasquazi, "Self-emergence of robust solitons in a microcavity," *Nature,* vol. 608, no. 7922, pp. 303-309, 2022/08/01, 2022.

[23] X. Xu, M. Tan, B. Corcoran, J. Wu, A. Boes, T. G. Nguyen, S. T. Chu, B. E. Little, D. G. Hicks, R. Morandotti, A. Mitchell, and D. J. Moss, "11 TOPS photonic convolutional accelerator for optical neural networks," *Nature,* vol. 589, no. 7840, pp. 44-51, 2021.

[24] B. Corcoran, M. Tan, X. Xu, A. Boes, J. Wu, T. G. Nguyen, S. T. Chu, B. E. Little, R. Morandotti, A. Mitchell, and D. J. Moss, "Ultra-dense optical data transmission over standard fibre with a single chip source," *Nature Communications,* vol. 11, no. 1, pp. 2568, 2020/05/22, 2020.

[25] J. Wu, Y. Yang, Y. Qu, X. Xu, Y. Liang, S. T. Chu, B. E. Little, R. Morandotti, B. Jia, and D. J. Moss, "Graphene Oxide Waveguide and Micro-Ring Resonator Polarizers," *Laser & Photonics Reviews,* vol. 13, no. 9, pp. 1900056, 2019/09/01, 2019.

[26] J. Wu, Y. Yang, Y. Qu, L. Jia, Y. Zhang, X. Xu, S. T. Chu, B. E. Little, R. Morandotti, B. Jia, and D. J. Moss, "2D Layered Graphene Oxide Films Integrated with Micro-Ring Resonators for Enhanced Nonlinear Optics," *Small,* vol. 16, no. 16, pp. 1906563, 2020/04/01, 2020.

[27] Y. Gao, W. Zhou, X. Sun, H. K. Tsang, and C. Shu, "Cavity-enhanced thermo-optic bistability and hysteresis in a graphene-on-Si3N4 ring resonator," *Optics Letters,* vol. 42, no. 10, pp. 1950-1953, 2017/05/15, 2017.

[28] G. Priem, P. Dumon, W. Bogaerts, D. V. Thourhout, G. Morthier, and R. Baets, "Optical bistability and pulsating behaviour in Silicon-On-Insulator ring resonator structures," *Optics Express,* vol. 13, no. 23, pp. 9623-9628, 2005/11/14, 2005.

[29] Y. Zhang, J. Wu, L. Jia, Y. Qu, Y. Yang, B. Jia, and D. J. Moss, "Graphene Oxide for Nonlinear Integrated Photonics," *Laser & Photonics Reviews,* vol. 17, no. 3, pp. 2200512, 2023/03/01, 2023.

[30] H. Arianfard, S. Juodkazis, D. J. Moss, and J. Wu, "Sagnac interference in integrated photonics," *Applied Physics Reviews,* vol. 10, no. 1, 2023.

[31] J. Wu, P. Cao, T. Pan, Y. Yang, C. Qiu, C. Tremblay, and Y. Su, "Compact on-chip 1 × 2 wavelength selective switch based on silicon microring resonator with nested pairs of subrings," *Photonics Research,* vol. 3, no. 1, pp. 9-14, 2015/02/01, 2015.

[32] G. Cocorullo, F. G. Della Corte, I. Rendina, and P. M. Sarro, "Thermo-optic effect exploitation in silicon microstructures," *Sensors and Actuators A: Physical,* vol. 71, no. 1, pp. 19-26, 1998/11/01/, 1998.

[33] A. Grieco, B. Slutsky, D. T. H. Tan, S. Zamek, M. P. Nezhad, and Y. Fainman, "Optical Bistability in a Silicon Waveguide Distributed Bragg Reflector Fabry–Pérot Resonator," *Journal of Lightwave Technology,* vol. 30, no. 14, pp. 2352-2355, 2012.

[34] M. Ferrera, D. Duchesne, L. Razzari, M. Peccianti, R. Morandotti, P. Cheben, S. Janz, D. X. Xu, B. E. Little, S. Chu, and D. J. Moss, "Low power four wave mixing in an integrated, micro-ring resonator with Q = 1.2 million," *Optics Express,* vol. 17, no. 16, pp. 14098-14103, 2009/08/03, 2009.

[35] J. Wu, T. Moein, X. Xu, G. Ren, A. Mitchell, and D. J. Moss, "Micro-ring resonator quality factor enhancement via an integrated Fabry-Perot cavity," *APL Photonics,* vol. 2, no. 5, 2017.

[36] H. Gao, Y. Jiang, Y. Cui, L. Zhang, J. Jia, and L. Jiang, "Investigation on the Thermo-Optic Coefficient of Silica Fiber Within a Wide Temperature Range," *Journal of Lightwave Technology,* vol. 36, no. 24, pp. 5881-5886, 2018.

[37] C. Horvath, D. Bachman, R. Indoe, and V. Van, "Photothermal nonlinearity and optical bistability in a graphene–silicon waveguide resonator," *Optics Letters,* vol. 38, no. 23, pp. 5036-5039, 2013.

[38] M. A. Foster, A. C. Turner, J. E. Sharping, B. S. Schmidt, M. Lipson, and A. L. Gaeta, "Broad-band optical parametric gain on a silicon photonic chip," *Nature,* vol. 441, no. 7096, pp. 960-963, 2006/06/01, 2006.

[39] J. Leuthold, C. Koos, and W. Freude, "Nonlinear silicon photonics," *Nature Photonics,* vol. 4, no. 8, pp. 535-544, 2010.

[40] J. Wu, H. Lin, D. J. Moss, K. P. Loh, and B. Jia, "Graphene oxide for photonics, electronics and optoelectronics," *Nature Reviews Chemistry,* vol. 7, no. 3, pp. 162-183, 2023/03/01, 2023.

[41] J. Wu, L. Jia, Y. Zhang, Y. Qu, B. Jia, and D. J. Moss, "Graphene Oxide for Integrated Photonics and Flat Optics," *Advanced Materials,* vol. 33, no. 3, pp. 2006415, 2021/01/01, 2021.

[42] J. Wu, B. Liu, J. Peng, J. Mao, X. Jiang, C. Qiu, C. Tremblay, and Y. Su, "On-Chip Tunable Second-Order Differential-Equation Solver Based on a Silicon Photonic Mode-Split Microresonator," *Journal of Lightwave Technology,* vol. 33, no. 17, pp. 3542-3549, 2015.

[43] K. Ikeda, R. E. Saperstein, N. Alic, and Y. Fainman, "Thermal and Kerr nonlinear properties of plasma-deposited silicon nitride/silicon dioxide waveguides," *Optics express,* vol. 16, no. 17, pp. 12987-12994, 2008.

[44] T. Gu, M. Yu, D.-L. Kwong, and C. W. Wong, "Molecular-absorption-induced thermal bistability in PECVD silicon nitride microring resonators," *Optics Express,* vol. 22, no. 15, pp. 18412-18420, 2014/07/28, 2014.

[45] L.-W. Luo, G. S. Wiederhecker, K. Preston, and M. Lipson, "Power insensitive silicon microring resonators," *Optics Letters,* vol. 37, no. 4, pp. 590-592, 2012/02/15, 2012.

[46] Y. Qu, J. Wu, Y. Zhang, L. Jia, Y. Liang, B. Jia, and D. J. Moss, "Analysis of Four-Wave Mixing in Silicon Nitride Waveguides Integrated With 2D Layered Graphene Oxide Films," *Journal of Lightwave Technology,* vol. 39, no. 9, pp. 2902-2910, 2021/05/01, 2021.

[47] Y. Zhang, J. Wu, Y. Qu, L. Jia, B. Jia, and D. J. Moss, "Design and Optimization of Four-Wave Mixing in Microring Resonators Integrated With 2D Graphene Oxide Films," *Journal of Lightwave Technology,* vol. 39, no. 20, pp. 6553-6562, 2021.

[48] W. Bogaerts, P. De Heyn, T. Van Vaerenbergh, K. De Vos, S. Kumar Selvaraja, T. Claes, P. Dumon, P. Bienstman, D. Van Thourhout, and R. Baets, "Silicon microring resonators," *Laser & Photonics Reviews,* vol. 6, no. 1, pp. 47-73, 2012.

[49]     Y. Yang, J. Wu, X. Xu, Y. Liang, S. T. Chu, B. E. Little, R. Morandotti, B. Jia, and D. J. Moss, "Invited Article: Enhanced four-wave mixing in waveguides integrated with graphene oxide," *APL Photonics,* vol. 3, no. 12, pp. 120803, 2018.

[50]     Q. Ma, T. Rossmann, and Z. Guo, "Temperature sensitivity of silica micro-resonators," *Journal of Physics D: Applied Physics,* vol. 41, no. 24, pp. 245111, 2008/12/03, 2008.

[51]     L. Jin, L. Di Lauro, A. Pasquazi, M. Peccianti, D. J. Moss, R. Morandotti, B. E. Little, and S. T. Chu, "Optical multi-stability in a nonlinear high-order microring resonator filter," *APL Photonics,* vol. 5, no. 5, 2020.

[52]     I. D. Rukhlenko, M. Premaratne, and G. P. Agrawal, "Analytical study of optical bistability in silicon ring resonators," *Optics Letters,* vol. 35, no. 1, pp. 55-57, 2010/01/01, 2010.

[53]     T. Gu, N. Petrone, J. F. McMillan, A. van der Zande, M. Yu, G.-Q. Lo, D.-L. Kwong, J. Hone, and C. W. Wong, "Regenerative oscillation and four-wave mixing in graphene optoelectronics," *Nature photonics,* vol. 6, no. 8, pp. 554-559, 2012.

[54]     Y. Zhang, L. Tao, D. Yi, J.-b. Xu, and H. K. Tsang, "Enhanced thermo-optic nonlinearities in a MoS2-on-silicon microring resonator," *Applied Physics Express,* vol. 13, no. 2, pp. 022004, 2020.

[55]     J. Wu, X. Xu, T. G. Nguyen, S. T. Chu, B. E. Little, R. Morandotti, A. Mitchell, and D. J. Moss, "RF Photonics: An Optical Microcombs' Perspective," *IEEE Journal of Selected Topics in Quantum Electronics,* vol. 24, no. 4, pp. 1-20, 2018.

[56]     S. Ghosh, I. Calizo, D. Teweldebrhan, E. P. Pokatilov, D. L. Nika, A. A. Balandin, W. Bao, F. Miao, and C. N. Lau, "Extremely high thermal conductivity of graphene: Prospects for thermal management applications in nanoelectronic circuits," *Applied Physics Letters,* vol. 92, no. 15, 2008.

[57]     P. Goli, S. Legedza, A. Dhar, R. Salgado, J. Renteria, and A. A. Balandin, "Graphene-enhanced hybrid phase change materials for thermal management of Li-ion batteries," *Journal of Power Sources,* vol. 248, pp. 37-43, 2014/02/15/, 2014.

[58]     J. D. Renteria, D. L. Nika, and A. A. Balandin, "Graphene Thermal Properties: Applications in Thermal Management and Energy Storage," *Applied Sciences,* 4, 2014].

[59]     A. Shakouri, "Nanoscale Thermal Transport and Microrefrigerators on a Chip," *Proceedings of the IEEE,* vol. 94, no. 8, pp. 1613-1638, 2006.

[60]     N. Mehra, L. Mu, T. Ji, X. Yang, J. Kong, J. Gu, and J. Zhu, "Thermal transport in polymeric materials and across composite interfaces," *Applied Materials Today,* vol. 12, pp. 92-130, 2018/09/01/, 2018.

[61]     C. Horvath, D. Bachman, R. Indoe, and V. Van, "Photothermal nonlinearity and optical bistability in a graphene-silicon waveguide resonator," *Optics Letters,* vol. 38, no. 23, pp. 5036-5039, 2013/12/01, 2013.

[62]     B. E. Little, S. T. Chu, P. P. Absil, J. V. Hryniewicz, F. G. Johnson, F. Seiferth, D. Gill, V. Van, O. King, and M. Trakalo, "Very high-order microring resonator filters for WDM applications," *IEEE Photonics Technology Letters,* vol. 16, no. 10, pp. 2263-2265, 2004.

[63]     B. Little, "A VLSI Photonics Platform," *Technical Digest.* p. ThD1.

[64]     A. Gyanathan, and Y.-C. Yeo, "Multi-level phase change memory devices with Ge2Sb2Te5 layers separated by a thermal insulating Ta2O5 barrier layer," *Journal of Applied Physics,* vol. 110, no. 12, 2011.

[65]     M. L. Grilli, D. Ristau, M. Dieckmann, and U. Willamowski, "Thermal conductivity of e-beam coatings," *Applied Physics A,* vol. 71, no. 1, pp. 71-76, 2000/07/01, 2000.

[66]     J. V. Campenhout, P. Rojo-Romeo, D. V. Thourhout, C. Seassal, P. Regreny, L. D. Cioccio, J. M. Fedeli, and R. Baets, "Thermal Characterization of Electrically Injected Thin-Film InGaAsP Microdisk Lasers on Si," *Journal of Lightwave Technology,* vol. 25, no. 6, pp. 1543-1548, 2007.

[67]  J. Komma, C. Schwarz, G. Hofmann, D. Heinert, and R. Nawrodt, "Thermo-optic coefficient of silicon at 1550 nm and cryogenic temperatures," *Applied Physics Letters,* vol. 101, no. 4, 2012.

[68]  G. Cocorullo, F. G. Della Corte, and I. Rendina, "Temperature dependence of the thermo-optic coefficient in crystalline silicon between room temperature and 550 K at the wavelength of 1523 nm," *Applied Physics Letters,* vol. 74, no. 22, pp. 3338-3340, 1999.

[69]  W.-C. Hsu, C. Zhen, and A. X. Wang, "Electrically Tunable High-Quality Factor Silicon Microring Resonator Gated by High Mobility Conductive Oxide," *ACS Photonics,* vol. 8, no. 7, pp. 1933-1936, 2021/07/21, 2021.

[70]  P. E. Barclay, K. Srinivasan, and O. Painter, "Nonlinear response of silicon photonic crystal microresonators excited via an integrated waveguide and fiber taper," *Optics Express,* vol. 13, no. 3, pp. 801-820, 2005/02/07, 2005.

[71]  D. Dai, Z. Wang, J. F. Bauters, M. C. Tien, M. J. R. Heck, D. J. Blumenthal, and J. E. Bowers, "Low-loss Si3N4 arrayed-waveguide grating (de)multiplexer using nano-core optical waveguides," *Optics Express,* vol. 19, no. 15, pp. 14130-14136, 2011/07/18, 2011.

[72]  X. Xu, J. Wu, M. Shoeiby, T. G. Nguyen, S. T. Chu, B. E. Little, R. Morandotti, A. Mitchell, and D. J. Moss, "Reconfigurable broadband microwave photonic intensity differentiator based on an integrated optical frequency comb source," APL Photonics, vol. 2, no. 9, 096104, Sep. 2017.

[73]  Xu, X., et al., Photonic microwave true time delays for phased array antennas using a 49 GHz FSR integrated micro-comb source, Photonics Research, 6, B30-B36 (2018).

[74]  X. Xu, M. Tan, J. Wu, R. Morandotti, A. Mitchell, and D. J. Moss, "Microcomb-based photonic RF signal processing", IEEE Photonics Technology Letters, vol. 31 no. 23 1854-1857, 2019.

[75]  Xu, et al., "Advanced adaptive photonic RF filters with 80 taps based on an integrated optical micro-comb source," Journal of Lightwave Technology, vol. 37, no. 4, pp. 1288-1295 (2019).

[76]  X. Xu, et al., "Photonic RF and microwave integrator with soliton crystal microcombs", IEEE Transactions on Circuits and Systems II: Express Briefs, vol. 67, no. 12, pp. 3582-3586, 2020.

[77]  X. Xu, et al., "High performance RF filters via bandwidth scaling with Kerr micro-combs," APL Photonics, vol. 4 (2) 026102. 2019.

[78]  M. Tan, et al., "Microwave and RF photonic fractional Hilbert transformer based on a 50 GHz Kerr micro-comb", Journal of Lightwave Technology, vol. 37, no. 24, pp. 6097 – 6104, 2019.

[79]  M. Tan, et al., "RF and microwave fractional differentiator based on photonics", IEEE Transactions on Circuits and Systems: Express Briefs, vol. 67, no.11, pp. 2767-2771, 2020.

[80]  M. Tan, et al., "Photonic RF arbitrary waveform generator based on a soliton crystal micro-comb source", Journal of Lightwave Technology, vol. 38, no. 22, pp. 6221-6226 (2020).

[81]  M. Tan, X. Xu, J. Wu, R. Morandotti, A. Mitchell, and D. J. Moss, "RF and microwave high bandwidth signal processing based on Kerr Micro-combs", Advances in Physics X, VOL. 6, NO. 1, 1838946 (2021). DOI:10.1080/23746149.2020.1838946.

[82]  X. Xu, et al., "Advanced RF and microwave functions based on an integrated optical frequency comb source," Opt. Express, vol. 26 (3) 2569 (2018).

[83]  M. Tan, X. Xu, J. Wu, B. Corcoran, A. Boes, T. G. Nguyen, S. T. Chu, B. E. Little, R.Morandotti, A. Lowery, A. Mitchell, and D. J. Moss, ""Highly Versatile Broadband RF Photonic Fractional Hilbert Transformer Based on a Kerr Soliton Crystal Microcomb", Journal of Lightwave Technology vol. 39 (24) 7581-7587 (2021).

[84]  Wu, J. et al. RF Photonics: An Optical Microcombs' Perspective. IEEE Journal of Selected Topics in Quantum Electronics Vol. 24, 6101020, 1-20 (2018).

[85]  T. G. Nguyen et al., "Integrated frequency comb source-based Hilbert transformer for

wideband microwave photonic phase analysis," Opt. Express, vol. 23, no. 17, pp. 22087-22097, Aug. 2015.

[86] X. Xu, et al., "Broadband RF channelizer based on an integrated optical frequency Kerr comb source," Journal of Lightwave Technology, vol. 36, no. 19, pp. 4519-4526, 2018.

[87] X. Xu, et al., "Continuously tunable orthogonally polarized RF optical single sideband generator based on micro-ring resonators," Journal of Optics, vol. 20, no. 11, 115701. 2018.

[88] X. Xu, et al., "Orthogonally polarized RF optical single sideband generation and dual-channel equalization based on an integrated microring resonator," Journal of Lightwave Technology, vol. 36, no. 20, pp. 4808-4818. 2018.

[89] X. Xu, et al., "Photonic RF phase-encoded signal generation with a microcomb source", J. Lightwave Technology, vol. 38, no. 7, 1722-1727, 2020.

[90] X. Xu, et al., Broadband microwave frequency conversion based on an integrated optical micro-comb source", Journal of Lightwave Technology, vol. 38 no. 2, pp. 332-338, 2020.

[91] M. Tan, et al., "Photonic RF and microwave filters based on 49GHz and 200GHz Kerr microcombs", Optics Comm. vol. 465,125563, Feb. 22. 2020.

[92] X. Xu, et al., "Broadband photonic RF channelizer with 90 channels based on a soliton crystal microcomb", Journal of Lightwave Technology, Vol. 38, no. 18, pp. 5116 – 5121 (2020).

[93] M. Tan et al, "Orthogonally polarized Photonic Radio Frequency single sideband generation with integrated micro-ring resonators", IOP Journal of Semiconductors, Vol. 42 (4), 041305 (2021).

[94] Mengxi Tan, X. Xu, J. Wu, T. G. Nguyen, S. T. Chu, B. E. Little, R. Morandotti, A. Mitchell, and David J. Moss, "Photonic Radio Frequency Channelizers based on Kerr Optical Micro-combs", IOP Journal of Semiconductors Vol. 42 (4), 041302 (2021). DOI:10.1088/1674-4926/42/4/041302.

[95] B. Corcoran, et al., "Ultra-dense optical data transmission over standard fiber with a single chip source", Nature Communications, vol. 11, Article:2568, 2020.

[96] X. Xu et al, "Photonic perceptron based on a Kerr microcomb for scalable high speed optical neural networks", Laser and Photonics Reviews, vol. 14, no. 8, 2000070 (2020).

[97] X. Xu, et al., "11 TOPs photonic convolutional accelerator for optical neural networks", Nature 589, 44-51 (2021).

[98] X. Xu et al., "Neuromorphic computing based on wavelength-division multiplexing", 28 IEEE Journal of Selected Topics in Quantum Electronics Vol. 29 Issue: 2, Article 7400112 (2023). DOI:10.1109/JSTQE.2022.3203159.

[99] Yang Sun, Jiayang Wu, Mengxi Tan, Xingyuan Xu, Yang Li, Roberto Morandotti, Arnan Mitchell, and David Moss, "Applications of optical micro-combs", Advances in Optics and Photonics 15 (1) 86-175 (2023).

[100] Yunping Bai, Xingyuan Xu,1, Mengxi Tan, Yang Sun, Yang Li, Jiayang Wu, Roberto Morandotti, Arnan Mitchell, Kun Xu, and David J. Moss, "Photonic multiplexing techniques for neuromorphic computing", Nanophotonics 12 (5): 795–817 (2023).

[101] Chawaphon Prayoonyong, Andreas Boes, Xingyuan Xu, Mengxi Tan, Sai T. Chu, Brent E. Little, Roberto Morandotti, Arnan Mitchell, David J. Moss, and Bill Corcoran, "Frequency comb distillation for optical superchannel transmission", Journal of Lightwave Technology 39 (23) 7383-7392 (2021).

[102] Mengxi Tan, Xingyuan Xu, Jiayang Wu, Bill Corcoran, Andreas Boes, Thach G. Nguyen, Sai T. Chu, Brent E. Little, Roberto Morandotti, Arnan Mitchell, and David J. Moss, "Integral order

photonic RF signal processors based on a soliton crystal micro-comb source", IOP Journal of Optics 23 (11) 125701 (2021).

[103] Yang Sun, Jiayang Wu, Yang Li, Xingyuan Xu, Guanghui Ren, Mengxi Tan, Sai Tak Chu, Brent E. Little, Roberto Morandotti, Arnan Mitchell, and David J. Moss, "Performance analysis of microcomb-based microwave photonic transversal signal processors with experimental errors", Journal of Lightwave Technology Vol. 41 Special Issue on Microwave Photonics (2023).

[104] Mengxi Tan, Xingyuan Xu, Andreas Boes, Bill Corcoran, Thach G. Nguyen, Sai T. Chu, Brent E. Little, Roberto Morandotti, Jiayang Wu, Arnan Mitchell, and David J. Moss, "Photonic signal processor for real-time video image processing at 17 Tb/s", Communications Engineering Vol. 2 (2023).

[105] Mengxi Tan, Xingyuan Xu, Jiayang Wu, Roberto Morandotti, Arnan Mitchell, and David J. Moss, "Photonic RF and microwave filters based on 49GHz and 200GHz Kerr microcombs", Optics Communications, 465, Article: 125563 (2020).

[106] Yang Sun, Jiayang Wu, Yang Li, Mengxi Tan, Xingyuan Xu, Sai Chu, Brent Little, Roberto Morandotti, Arnan Mitchell, and David J. Moss, "Quantifying the Accuracy of Microcomb-based Photonic RF Transversal Signal Processors", IEEE Journal of Selected Topics in Quantum Electronics 29 no. 6, pp. 1-17, Art no. 7500317 (2023)..

[107] Kues, M. et al. "Quantum optical microcombs", Nature Photonics 13, (3) 170-179 (2019).

[108] C.Reimer, L. Caspani, M. Clerici, et al., "Integrated frequency comb source of heralded single photons," Optics Express, vol. 22, no. 6, pp. 6535-6546, 2014.

[109] C.Reimer, et al., "Cross-polarized photon-pair generation and bi-chromatically pumped optical parametric oscillation on a chip", Nature Communications, vol. 6, Article 8236, 2015.

[110] L. Caspani, C. Reimer, M. Kues, et al., "Multifrequency sources of quantum correlated photon pairs on-chip: a path toward integrated Quantum Frequency Combs," Nanophotonics, vol. 5, no. 2, pp. 351-362, 2016.

[111] C. Reimer et al., "Generation of multiphoton entangled quantum states by means of integrated frequency combs," Science, vol. 351, no. 6278, pp. 1176-1180, 2016.

[112] M. Kues, et al., "On-chip generation of high-dimensional entangled quantum states and their coherent control", Nature, vol. 546, no. 7660, pp. 622-626, 2017.

[113] P. Roztocki et al., "Practical system for the generation of pulsed quantum frequency combs," Optics Express, vol. 25, no. 16, pp. 18940-18949, 2017.

[114] Y. Zhang, et al., "Induced photon correlations through superposition of two four-wave mixing processes in integrated cavities", Laser and Photonics Reviews, vol. 14, no. 7, pp. 2000128, 2020.

[115] C. Reimer, et al., "High-dimensional one-way quantum processing implemented on d-level cluster states", Nature Physics, vol. 15, no.2, pp. 148–153, 2019.

[116] P.Roztocki et al., "Complex quantum state generation and coherent control based on integrated frequency combs", Journal of Lightwave Technology 37 (2) 338-347 (2019).

[117] S. Sciara et al., "Generation and Processing of Complex Photon States with Quantum Frequency Combs", IEEE Photonics Technology Letters 31 (23) 1862-1865 (2019).

[118] Stefania Sciara, Piotr Roztocki, Bennet Fisher, Christian Reimer, Luis Romero Cortez, William J. Munro, David J. Moss, Alfonso C. Cino, Lucia Caspani, Michael Kues, J. Azana, and Roberto Morandotti, "Scalable and effective multilevel entangled photon states: A promising tool to boost quantum technologies", Nanophotonics 10 (18), 4447–4465 (2021).

[119] L. Caspani, C. Reimer, M. Kues, et al., "Multifrequency sources of quantum correlated photon

pairs on-chip: a path toward integrated Quantum Frequency Combs," Nanophotonics, vol. 5, no. 2, pp. 351-362, 2016.

[120] Moss, "Enhanced supercontinuum generated in SiN waveguides coated with GO films", Advanced Materials Technologies 8 (1) 2201796 (2023). DOI: 10.1002/admt.202201796.

[121] Yuning Zhang, Jiayang Wu, Linnan Jia, Yang Qu, Baohua Jia, and David J. Moss, "Graphene oxide for nonlinear integrated photonics", Laser and Photonics Reviews 17 2200512 (2023).

[122] Jiayang Wu, H. Lin, D. J. Moss, T.K. Loh, Baohua Jia, "Graphene oxide: new opportunities for electronics, photonics, and optoelectronics", Nature Reviews Chemistry 7 (3) 162–183 (2023).

[123] Yang Qu, Jiayang Wu, Yuning Zhang, Yunyi Yang, Linnan Jia, Baohua Jia, and David J. Moss, "Photo thermal tuning in GO-coated integrated waveguides", Micromachines 13 1194 (2022).

[124] Yuning Zhang, Jiayang Wu, Yunyi Yang, Yang Qu, Houssein El Dirani, Romain Crochemore, Corrado Sciancalepore, Pierre Demongodin, Christian Grillet, Christelle Monat, Baohua Jia, and David J. Moss, "Enhanced self-phase modulation in silicon nitride waveguides integrated with 2D graphene oxide films", IEEE Journal of Selected Topics in Quantum Electronics 29 (1) 5100413 (2023).

[125] Yuning Zhang, Jiayang Wu, Yunyi Yang, Yang Qu, Linnan Jia, Baohua Jia, and David J. Moss, "Enhanced spectral broadening of femtosecond optical pulses in silicon nanowires integrated with 2D graphene oxide films", Micromachines 13 756 (2022).

[126] Linnan Jia, Jiayang Wu, Yuning Zhang, Yang Qu, Baohua Jia, Zhigang Chen, and David J. Moss, "Fabrication Technologies for the On-Chip Integration of 2D Materials", Small: Methods 6, 2101435 (2022).

[127] Yuning Zhang, Jiayang Wu, Yang Qu, Linnan Jia, Baohua Jia, and David J. Moss, "Design and optimization of four-wave mixing in microring resonators integrated with 2D graphene oxide films", Journal of Lightwave Technology 39 (20) 6553-6562 (2021).

[128] Yuning Zhang, Jiayang Wu, Yang Qu, Linnan Jia, Baohua Jia, and David J. Moss, "Optimizing the Kerr nonlinear optical performance of silicon waveguides integrated with 2D graphene oxide films", Journal of Lightwave Technology 39 (14) 4671-4683 (2021).

[129] Yang Qu, Jiayang Wu, Yuning Zhang, Yao Liang, Baohua Jia, and David J. Moss, "Analysis of four-wave mixing in silicon nitride waveguides integrated with 2D layered graphene oxide films", Journal of Lightwave Technology 39 (9) 2902-2910 (2021).

[130] Jiayang Wu, Linnan Jia, Yuning Zhang, Yang Qu, Baohua Jia, and David J. Moss," Graphene oxide: versatile films for flat optics to nonlinear photonic chips", Advanced Materials 33 (3) 2006415, pp.1-29 (2021).

[131] Y. Qu, J. Wu, Y. Zhang, L. Jia, Y. Yang, X. Xu, S. T. Chu, B. E. Little, R. Morandotti, B. Jia, and D. J. Moss, "Graphene oxide for enhanced optical nonlinear performance in CMOS compatible integrated devices", Paper No. 11688-30, PW21O-OE109-36, 2D Photonic Materials and Devices IV, SPIE Photonics West, San Francisco CA March 6-11 (2021).

[132] Yang Qu, Jiayang Wu, Yunyi Yang, Yuning Zhang, Yao Liang, Houssein El Dirani, Romain Crochemore, Pierre Demongodin, Corrado Sciancalepore, Christian Grillet, Christelle Monat, Baohua Jia, and David J. Moss, "Enhanced nonlinear four-wave mixing in silicon nitride waveguides integrated with 2D layered graphene oxide films", Advanced Optical Materials vol. 8 (21) 2001048 (2020).

[133] Yuning Zhang, Yang Qu, Jiayang Wu, Linnan Jia, Yunyi Yang, Xingyuan Xu, Baohua Jia, and David J. Moss, "Enhanced Kerr nonlinearity and nonlinear figure of merit in silicon nanowires integrated with 2D graphene oxide films", ACS Applied Materials and Interfaces vol. 12 (29)

33094−33103 June 29 (2020).

[134] Jiayang Wu, Yunyi Yang, Yang Qu, Yuning Zhang, Linnan Jia, Xingyuan Xu, Sai T. Chu, Brent E. Little, Roberto Morandotti, Baohua Jia,* and David J. Moss*, "Enhanced nonlinear four-wave mixing in microring resonators integrated with layered graphene oxide films", Small vol. 16 (16) 1906563 April 23 (2020).

[135] Jiayang Wu, Yunyi Yang, Yang Qu, Xingyuan Xu, Yao Liang, Sai T. Chu, Brent E. Little, Roberto Morandotti, Baohua Jia, and David J. Moss, "Graphene oxide waveguide polarizers and polarization selective micro-ring resonators", Paper 11282-29, SPIE Photonics West, San Francisco, CA, 4 - 7 February (2020).

[136] Jiayang Wu, Yunyi Yang, Yang Qu, Xingyuan Xu, Yao Liang, Sai T. Chu, Brent E. Little, Roberto Morandotti, Baohua Jia, and David J. Moss, "Graphene oxide waveguide polarizers and polarization selective micro-ring resonators", Laser and Photonics Reviews vol. 13 (9) 1900056 (2019).

[137] Yunyi Yang, Jiayang Wu, Xingyuan Xu, Sai T. Chu, Brent E. Little, Roberto Morandotti, Baohua Jia, and David J. Moss, "Enhanced four-wave mixing in graphene oxide coated waveguides", Applied Physics Letters Photonics vol. 3 120803 (2018).

[138] Linnan Jia, Yang Qu, Jiayang Wu, Yuning Zhang, Yunyi Yang, Baohua Jia, and David J. Moss, "Third-order optical nonlinearities of 2D materials at telecommunications wavelengths", Micromachines (MDPI), 14, 307 (2023).

[139] A. Pasquazi, et al., "Sub-picosecond phase-sensitive optical pulse characterization on a chip", Nature Photonics, vol. 5, no. 10, pp. 618-623 (2011).

[140] Bao, C., et al., Direct soliton generation in microresonators, Opt. Lett, 42, 2519 (2017).

[141] M.Ferrera et al., "CMOS compatible integrated all-optical RF spectrum analyzer", Optics Express, vol. 22, no. 18, 21488 - 21498 (2014).

[142] M. Kues, et al., "Passively modelocked laser with an ultra-narrow spectral width", Nature Photonics, vol. 11, no. 3, pp. 159, 2017.

[143] L. Razzari, et al., "CMOS-compatible integrated optical hyper-parametric oscillator," Nature Photonics, vol. 4, no. 1, pp. 41-45, 2010.

[144] M. Ferrera, et al., "Low-power continuous-wave nonlinear optics in doped silica glass integrated waveguide structures," Nature Photonics, vol. 2, no. 12, pp. 737-740, 2008.

[145] M.Ferrera et al."On-Chip ultra-fast 1st and 2nd order CMOS compatible all-optical integration", Opt. Express, vol. 19, (23)pp. 23153-23161 (2011).

[146] D. Duchesne, M. Peccianti, M. R. E. Lamont, et al., "Supercontinuum generation in a high index doped silica glass spiral waveguide," Optics Express, vol. 18, no, 2, pp. 923-930, 2010.

[147] H Bao, L Olivieri, M Rowley, ST Chu, BE Little, R Morandotti, DJ Moss, … "Turing patterns in a fiber laser with a nested microresonator: Robust and controllable microcomb generation", Physical Review Research 2 (2), 023395 (2020).

[148] M. Ferrera, et al., "On-chip CMOS-compatible all-optical integrator", Nature Communications, vol. 1, Article 29, 2010.

[149] A. Pasquazi, et al., "All-optical wavelength conversion in an integrated ring resonator," Optics Express, vol. 18, no. 4, pp. 3858-3863, 2010.

[150] A.Pasquazi, Y. Park, J. Azana, et al., "Efficient wavelength conversion and net parametric gain via Four Wave Mixing in a high index doped silica waveguide," Optics Express, vol. 18, no. 8, pp. 7634-7641, 2010.

[151] M. Peccianti, M. Ferrera, L. Razzari, et al., "Subpicosecond optical pulse compression via an

integrated nonlinear chirper," Optics Express, vol. 18, no. 8, pp. 7625-7633, 2010.

[152] Little, B. E. et al., "Very high-order microring resonator filters for WDM applications", IEEE Photonics Technol. Lett. 16, 2263–2265 (2004).

[153] M. Ferrera et al., "Low Power CW Parametric Mixing in a Low Dispersion High Index Doped Silica Glass Micro-Ring Resonator with Q-factor > 1 Million", Optics Express, vol.17, no. 16, pp. 14098–14103 (2009).

[154] M. Peccianti, et al., "Demonstration of an ultrafast nonlinear microcavity modelocked laser", Nature Communications, vol. 3, pp. 765, 2012.

[155] A.Pasquazi, et al., "Self-locked optical parametric oscillation in a CMOS compatible microring resonator: a route to robust optical frequency comb generation on a chip," Optics Express, vol. 21, no. 11, pp. 13333-13341, 2013.

[156] A.Pasquazi, et al., "Stable, dual mode, high repetition rate mode-locked laser based on a microring resonator," Optics Express, vol. 20, no. 24, pp. 27355-27362, 2012.

[157] Pasquazi, A. et al. Micro-combs: a novel generation of optical sources. Physics Reports 729, 1-81 (2018).

[158] Moss, D. J. et al., "New CMOS-compatible platforms based on silicon nitride and Hydex for nonlinear optics", Nature photonics 7, 597 (2013).

[159] H. Bao, et al., Laser cavity-soliton microcombs, Nature Photonics, vol. 13, no. 6, pp. 384-389, Jun. 2019.

[160] Antonio Cutrona, Maxwell Rowley, Debayan Das, Luana Olivieri, Luke Peters, Sai T. Chu, Brent L. Little, Roberto Morandotti, David J. Moss, Juan Sebastian Totero Gongora, Marco Peccianti, Alessia Pasquazi, "High Conversion Efficiency in Laser Cavity-Soliton Microcombs", Optics Express Vol. 30, Issue 22, pp. 39816-39825 (2022). https://doi.org/10.1364/OE.470376.

[161] M.Rowley, P.Hanzard, A.Cutrona, H.Bao, S.Chu, B.Little, R.Morandotti, D. J. Moss, G. Oppo, J. Gongora, M. Peccianti and A. Pasquazi, "Self-emergence of robust solitons in a micro-cavity", Nature 608 (7922) 303–309 (2022).

[162] A. Cutrona, M. Rowley, A. Bendahmane, V. Cecconi,L. Peters, L. Olivieri, B. E. Little, S. T. Chu, S. Stivala, R. Morandotti, D. J. Moss, J. S. Totero-Gongora, M. Peccianti, A. Pasquazi, "Nonlocal bonding of a soliton and a blue-detuned state in a microcomb laser", Nature Communications Physics 6 (2023).

[163] A. Cutrona, M. Rowley, A. Bendahmane, V. Cecconi,L. Peters, L. Olivieri, B. E. Little, S. T. Chu, S. Stivala, R. Morandotti, D. J. Moss, J. S. Totero-Gongora, M. Peccianti, A. Pasquazi, "Stability Properties of Laser Cavity-Solitons for Metrological Applications", Applied Physics Letters 122 (12) 121104 (2023); doi: 10.1063/5.0134147.

[164] Hamed Arianfard, Saulius Juodkazis, David J. Moss, and Jiayang Wu, "Sagnac interference in integrated photonics", Applied Physics Reviews vol. 10 (1) 011309 (2023).

[165] Hamed Arianfard, Jiayang Wu, Saulius Juodkazis, and David J. Moss, "Optical analogs of Rabi splitting in integrated waveguide-coupled resonators", Advanced Physics Research 2 (2023).

[166] Hamed Arianfard, Jiayang Wu, Saulius Juodkazis, and David J. Moss, "Spectral shaping based on optical waveguides with advanced Sagnac loop reflectors", Paper No. PW22O-OE201-20, SPIE-Opto, Integrated Optics: Devices, Materials, and Technologies XXVI, SPIE Photonics West, San Francisco CA January 22 - 27 (2022).

[167] Hamed Arianfard, Jiayang Wu, Saulius Juodkazis, David J. Moss, "Spectral Shaping Based on Integrated Coupled Sagnac Loop Reflectors Formed by a Self-Coupled Wire Waveguide", IEEE Photonics Technology Letters vol. 33 (13) 680-683 (2021).

[168] 5. Hamed Arianfard, Jiayang Wu, Saulius Juodkazis and David J. Moss, "Three Waveguide Coupled Sagnac Loop Reflectors for Advanced Spectral Engineering", Journal of Lightwave Technology vol. 39 (11) 3478-3487 (2021).

[169] Hamed Arianfard, Jiayang Wu, Saulius Juodkazis and David J. Moss, "Advanced Multi-Functional Integrated Photonic Filters based on Coupled Sagnac Loop Reflectors", Journal of Lightwave Technology vol. 39 Issue: 5, pp.1400-1408 (2021). DOI:10.1109/JLT.2020.3037559.

[170] Hamed Arianfard, Jiayang Wu, Saulius Juodkazis and David J. Moss, "Advanced multi-functional integrated photonic filters based on coupled Sagnac loop reflectors", Paper 11691-4, PW21O-OE203-44, Silicon Photonics XVI, SPIE Photonics West, San Francisco CA March 6-11 (2021).

[171] Jiayang Wu, Tania Moein, Xingyuan Xu, and David J. Moss, "Advanced photonic filters via cascaded Sagnac loop reflector resonators in silicon-on-insulator integrated nanowires", Applied Physics Letters Photonics vol. 3 046102 (2018). DOI:/10.1063/1.5025833

[172] Jiayang Wu, Tania Moein, Xingyuan Xu, Guanghui Ren, Arnan Mitchell, and David J. Moss, "Micro-ring resonator quality factor enhancement via an integrated Fabry-Perot cavity", Applied Physics Letters Photonics vol. 2 056103 (2017). doi: 10.1063/1.4981392.

[173] Linnan Jia, Dandan Cui, Jiayang Wu, Haifeng Feng, Tieshan Yang, Yunyi Yang, Yi Du, Weichang Hao, Baohua Jia, David J. Moss, "BiOBr nanoflakes with strong nonlinear optical properties towards hybrid integrated photonic devices", Applied Physics Letters Photonics vol. 4 090802 (2019).

[174] Linnan Jia, Jiayang Wu, Yunyi Yang, Yi Du, Baohua Jia, David J. Moss, "Large Third-Order Optical Kerr Nonlinearity in Nanometer-Thick PdSe2 2D Dichalcogenide Films: Implications for Nonlinear Photonic Devices", ACS Applied Nano Materials vol. 3 (7) 6876–6883 (2020).

[175] E.D Ghahramani, DJ Moss, JE Sipe, "Full-band-structure calculation of first-, second-, and third-harmonic optical response coefficients of ZnSe, ZnTe, and CdTe", Physical Review B 43 (12), 9700 (1991).

[176] C Grillet, C Smith, D Freeman, S Madden, B Luther-Davies, EC Magi, … "Efficient coupling to chalcogenide glass photonic crystal waveguides via silica optical fiber nanowires", Optics Express vol. 14 (3), 1070-1078 (2006).

[177] S Tomljenovic-Hanic, MJ Steel, CM de Sterke, DJ Moss, "High-Q cavities in photosensitive photonic crystals" Optics Letters vol. 32 (5), 542-544 (2007).

[178] M Ferrera et al., "On-Chip ultra-fast 1st and 2nd order CMOS compatible all-optical integration", Optics Express vol. 19 (23), 23153-23161 (2011).

[179] VG Ta'eed et al., "Error free all optical wavelength conversion in highly nonlinear As-Se chalcogenide glass fiber", Optics Express vol. 14 (22), 10371-10376 (2006).

[180] M Rochette, L Fu, V Ta'eed, DJ Moss, BJ Eggleton, "2R optical regeneration: an all-optical solution for BER improvement", IEEE Journal of Selected Topics in Quantum Electronics vol. 12 (4), 736-744 (2006).

[181] TD Vo, et al., "Silicon-chip-based real-time dispersion monitoring for 640 Gbit/s DPSK signals", Journal of Lightwave Technology vol. 29 (12), 1790-1796 (2011).

[182] Di Jin, Jiayang Wu, Sian Ren, Junkai Hu, Duan Huang, and David J. Moss, "Modeling of complex integrated photonic resonators using scattering matrix method", Photonics, Vol. 11, 1107 (2024). https://doi.org/10.3390/photonics11121107.

[183] Yonghang Sun, James Salamy, Caitlin E. Murry, Brent E. Little, Sai T. Chu, Roberto Morandotti, Arnan Mitchell, David J. Moss, Bill Corcoran, "Enhancing laser temperature stability by passive

self-injection locking to a micro-ring resonator", Optics Express Vol. 32 (13) 23841-23855 (2024). https://doi.org/10.1364/OE.515269.

[184] C. Khallouf, V. T. Hoang, G. Fanjoux, B. Little, S. T. Chu, D. J. Moss, R. Morandotti, J. M. Dudley, B. Wetzel, and T. Sylvestre, "Raman scattering and supercontinuum generation in high-index doped silica chip waveguides", Nonlinear Optics and its Applications, edited by John M. Dudley, Anna C. Peacock, Birgit Stiller, Giovanna Tissoni, SPIE Vol. 13004, 130040I (2024). doi: 10.1117/12.3021965

[185] Yang Li, Yang Sun, Jiayang Wu, Guanghui Ren, Roberto Morandotti, Xingyuan Xu, Mengxi Tan, Arnan Mitchell, and David J. Moss, "Performance analysis of microwave photonic spectral filters based on optical microcombs", Advanced Physics Research, Vol. 3 (9) 2400084 (2024). DOI:10.1002/apxr.202400084.

[186] Di Jin, Jiayang Wu, Junkai Hu, Wenbo Liu, Yuning Zhang, Yunyi Yang, Linnan Jia, Duan Huang, Baohua Jia, and David J. Moss, "Silicon photonic waveguide and microring resonator polarizers incorporating 2D graphene oxide films", Applied Physics Letters, Vol. 125, 053101 (2024). doi: 10.1063/5.0221793.

[187] Jiayang Wu, Yuning Zhang, Junkai Hu, Yunyi Yang, Di Jin, Wenbo Liu, Duan Huang, Baohua Jia, David J. Moss, "Novel functionality with 2D graphene oxide films integrated on silicon photonic chips", Advanced Materials, Vol. 36, 2403659 (2024). DOI: 10.1002/adma.202403659.

[188] Yuning Zhang, Jiayang Wu, Linnan Jia, Di Jin, Baohua Jia, Xiaoyong Hu, David Moss, Qihuang Gong, "Advanced optical polarizers based on 2D materials", npj Nanophotonics, Vol. 1 (2024). DOI: 10.1038/s44310-024-00028-3.

[189] Junkai Hu, Jiayang Wu, Wenbo Liu, Di Jin, Houssein El Dirani, Sébastien Kerdiles, Corrado Sciancalepore, Pierre Demongodin, Christian Grillet, Christelle Monat, Duan Huang, Baohua Jia, and David J. Moss, "2D graphene oxide: a versatile thermo-optic material", Advanced Functional Materials, Vol. 34, 2406799 (2024). DOI: 10.1002/adfm.202406799.

[190] Di Jin, Wenbo Liu, Linnan Jia, Junkai Hu, Duan Huang, Jiayang Wu, Baohua Jia, and David J. Moss, "Thickness and Wavelength Dependent Nonlinear Optical Absorption in 2D Layered MXene Films", Small Science Vol. 4, 2400179 (2024). DOI:10.1002/smsc202400179.

[191] Andrew Cooper, Luana Olivieri, Antonio Cutrona, Debayan Das, Luke Peters, Sai Tak Chu, Brent Little, Roberto Morandotti, David J Moss, Marco Peccianti, and Alessia Pasquazi, "Parametric interaction of laser cavity-solitons with an external CW pump", Optics Express Vol. 32 (12), 21783-21794 (2024).

[192] Weiwei Han, Zhihui Liu, Yifu Xu, Mengxi Tan, Chaoran Huang, Jiayang Wu, Kun Xu, David J. Moss, and Xingyuan Xu, "Photonic RF Channelization Based on Microcombs", Special Issue on Microcombs IEEE Journal of Selected Topics in Quantum Electronics Vol. 30 (5) 7600417 (2024). DOI:10.1109/JSTQE.2024.3398419.

[193] Y. Li, Y. Sun, J. Wu, G. Ren, X. Xu, M. Tan, S. Chu, B. Little, R. Morandotti, A. Mitchell, and D. J. Moss, "Feedback control in micro-comb-based microwave photonic transversal filter systems", IEEE Journal of Selected Topics in Quantum Electronics Vol. 30 (5) 2900117 (2024). DOI: 10.1109/JSTQE.2024.3377249.

[194] Weiwei Han, Zhihui Liu, Yifu Xu, Mengxi Tan, Yuhua Li, Xiaotian Zhu, Yanni Ou, Feifei Yin, Roberto Morandotti, Brent E. Little, Sai Tak Chu, Xingyuan Xu, David J. Moss, and Kun Xu, "Dual-polarization RF Channelizer Based on Microcombs", Optics Express Vol. 32, No. 7, 11281-11295  / 25 Mar 2024 / (2024).   DOI: 10.1364/OE.519235.

[195]   Aadhi A. Rahim, Imtiaz Alamgir, Luigi Di Lauro, Bennet Fischer, Nicolas Perron, Pavel Dmitriev, Celine Mazoukh, Piotr Roztocki, Cristina Rimoldi, Mario Chemnitz, Armaghan Eshaghi, Evgeny A. Viktorov, Anton V. Kovalev, Brent E. Little, Sai T. Chu, David J. Moss, and Roberto Morandotti, "Mode-locked laser with multiple timescales in a microresonator-based nested cavity", APL Photonics Vol. 9, 031302 (2024). DOI:10.1063/5.0174697.

[196]   C. Mazoukh, L. Di Lauro, I. Alamgir1 B. Fischer, A. Aadhi, A. Eshaghi, B. E. Little, S. T. Chu, D. J. Moss, and R. Morandotti, "Genetic algorithm-enhanced microcomb state generation", Special Issue Microresontaor Frequency Combs - New Horizons, Nature Communications Physics, Vol. 7, Article: 81 (2024) (2024). DOI: 10.1038/s42005-024-01558-0.

[197]   Y. Zhang, J. Wu, Y. Yang,Y. Qu, L. Jia, C. Grillet, C. Monat, B. Jia, and D.J. Moss, "Graphene oxide for enhanced nonlinear optics in integrated photonic chips", Paper 12888-16, Conference OE109, 2D Photonic Materials and Devices VII, Chair(s): Arka Majmdar; Carlos M. Torres Jr.; Hui Deng, SPIE Photonics West, San Francisco CA, January 27 – February 1 (2024). Proceedings Volume 12888, 2D Photonic Materials and Devices VII; 1288805 (2024). https://doi.org/10.1117/12.3005069

[198]   Zhang Y, Wu J, Qu Y, Jia L, Jia B, D.J. Moss, "Graphene oxide-based waveguides for enhanced self-phase modulation", Annals of Mathematics and Physics Vol. 5 (2) 103-106 (2022). DOI:10.17352/amp.000048

[199]   Bill Corcoran, Arnan Mitchell, Roberto Morandotti, Leif K. Oxenlowe, and David J. Moss, "Microcombs for Optical Communications", **Nature Photonics,** Vol. 19 (2025).